\documentclass[structabstract]{aa}

\usepackage{graphicx}
\usepackage{txfonts}
\usepackage[english]{babel}
\usepackage{url}
\usepackage{lscape}

\usepackage{natbib}
\bibpunct{(}{)}{;}{a}{}{,}

\begin{document}

\title{Methanol maps of low-mass protostellar systems}
\subtitle{I. The Serpens Molecular Core}

\author{L.E. Kristensen\inst{1} \and E.F. van Dishoeck\inst{1,2} \and
  T.A. van Kempen\inst{3} \and H.M. Cuppen\inst{1} \and
  C. Brinch\inst{1} \and J.K. J{\o}rgensen\inst{4} \and
  M.R. Hogerheijde\inst{1}}

\institute{Leiden Observatory, Leiden University, PO Box 9513, 2300 RA
  Leiden, The Netherlands\\
  \email{kristensen@strw.leidenuniv.nl} \and 
  Max Planck Institut f{\"u}r Extraterrestrische Physik (MPE),
  Giessenbachstrasse 1, 85748 Garching, Germany \and
  Harvard-Smithsonian Center for Astrophysics, 60 Garden Street, MS
  78, Cambridge, MA 02138, USA \and
  Centre for Star and Planet Formation, Natural History Museum of
  Denmark, {\O}ster Voldgade 5-7, DK-1350 Copenhagen K., Denmark
}

\date{\today}

\abstract{Methanol has a rich rotational spectrum providing a large
  number of transitions at sub-millimetre wavelengths from a range of
  energy levels in one single
  telescope setting, thus making it a good tracer of physical
  conditions in star-forming regions. Furthermore, it is formed
  exclusively on grain surfaces and is therefore a clean tracer of
  surface chemistry.}
{Determining the physical and chemical structure of low-mass,
  young stellar objects, in particular the abundance structure of
  CH$_3$OH, to investigate where and how CH$_3$OH forms and how it is
  eventually released back to the gas phase.}
{Observations of the Serpens Molecular Core have been performed at the
  James Clerk Maxwell Telescope using the array receiver, Harp-B. Maps
  over a 4\farcm5$\times$5\farcm4 region were made in a frequency
  window around 338 GHz, covering the 7$_K$--6$_K$ transitions of
  methanol. Data are compared with physical models of each
  source based on existing sub-millimetre continuum data.}
{Methanol emission is extended over each source, following the
  column density of H$_2$ but showing up also particularly strongly
  around outflows. The rotational temperature is low, 15--20~K, and
  does not vary with position within each source. None of the Serpens
  Class 0 sources show the high-$K$ lines seen in several other Class 0
  sources. 
  The abundance is typically 10$^{-9}$ -- 10$^{-8}$ with respect to
  H$_2$ in the outer envelope,
  whereas ``jumps'' by factors of up to 10$^{2}$--10$^{3}$ inside the
  region where the dust temperature exceeds 100 K are not excluded. A
  factor of up to $\sim$ 10$^3$ enhancement is seen in
  outflow gas, consistent with previous studies. In one object, 
  SMM4, the ice abundance has been measured to be $\sim$~3~$\times$
  10$^{-5}$ with respect to H$_2$ in the outer envelope, i.e., a
  factor of 10$^3$ larger than the gas-phase abundance. Comparison with
  C$^{18}$O $J$=3--2 emission shows that strong CO depletion leads to a high
  gas-phase abundance of CH$_3$OH not just for the Serpens sources, but
  also for a larger sample of deeply embedded protostars.}
{The observations illustrate the large-scale, low-level desorption of
  CH$_3$OH from dust grains, extending out to and beyond 7500~AU from
  each source, a scenario which is consistent with non-thermal
  \mbox{(photo-)desorption} from the ice. The observations also
  illustrate the usefulness of CH$_3$OH as a tracer of energetic input
  in the form of outflows, where methanol is sputtered from the grain
  surfaces. Finally, the observations provide further
  evidence of CH$_3$OH formation through CO hydrogenation proceeding
  on grain surfaces in low-mass envelopes.
}


\maketitle

\section{Introduction}

A long-standing goal of astrochemistry has been to determine the
physical and chemical conditions prevailing in star-forming
regions \citep[e.g.,][]{vandishoeck98}. In this respect, different
molecules act as tracers of different physical components, all
depending on their formation history, their abundances, their chemical
properties, etc. To effectively trace physical conditions such as
density and temperature over the large range of values found in
star-forming regions over the time-scale of star-formation, it is of
great importance to have as many independent tracers as
possible. Methanol (CH$_3$OH), with its rich rotational
spectrum, is an excellent candidate tracing both temperature, density,
grain surface formation and energy injection simultaneously during all
phases of the early stages of stellar evolution.

CH$_3$OH is a slightly asymmetric top molecule with numerous
rotational transitions observable at millimetre- and sub-millimetre
wavelengths. Because of the large number of transitions observable in
a single frequency window it is possible to obtain a coherent data set
very efficiently, making methanol a very suitable tracer of physical
conditions, in particular in low-mass star forming regions where the
emission is optically thin. Moreover, since the molecule is a slightly
asymmetric top molecule, it traces very efficiently both density and
temperature \citep[e.g.,][]{maret05, jorgensen05, leurini07}.

Methanol forms exclusively on ice-covered dust grain surfaces
primarily through hydrogenation of CO \citep{watanabe02, fuchs09}. Observations
of interstellar ices show that methanol is indeed a prominent ice
component, with abundances of up to almost 30\% with respect to
solid-state H$_2$O or a few $\times$ 10$^{-5}$ with respect to gas-phase
H$_2$ \citep[e.g.,][]{dartois99, gibb04, pontoppidan04}. In contrast,
pure gas phase reactions produce negligible CH$_3$OH abundances of
less than 10$^{-10}$ \citep{garrod06}. The question that naturally
arises is how methanol desorbs from the surface of a dust grain to
be observed in the gas phase. Is it through thermal heating of the
entire grain, or is it through non-thermal desorption, where cosmic
rays, UV-photons or exothermic reactions provide local heating of the
grain? The former mechanism is at play close to young
stellar objects (YSOs), in the inner-most part of the molecular
envelope where the gas temperature exceeds 100~K
\citep{vandishoeck95, ceccarelli00, vandertak00, schoier02, maret05,
  jorgensen05} and in outflows where hot gas
sputters the icy mantles \citep[e.g.][]{bachiller95, bachiller97}. This
mechanism allows for an effective methanol enrichment of the
environment and abundances are typically in the range of 10$^{-7}$ to
10$^{-6}$. The non-thermal
mechanism dominates in cold, dark clouds and the outer parts of
molecular envelopes \citep{hasegawa93, herbst06, garrod07, oberg09a}. Here,
reported abundances typically have values of 10$^{-10}$ to
10$^{-8}$. Hence, methanol also acts as a tracer of energetic processes
in star-forming regions.

So far most studies have concentrated on spectra at a single position
or at most a few around them. Recently, large-scale mapping of weak
molecular lines has become very efficient with the advent of array
receivers such as the 16 pixel Harp-B receiver on the
James Clerk Maxwell Telescope (JCMT). This
allows for a direct study of the entire protostellar system (primarily
envelope and outflow) on scales of several arcminutes at
15\arcsec\ resolution and for determination of density, temperature
and energy input into the system. These observations will eventually
be compared directly to observations of another important grain
surface product, H$_2$O, to be done with the Herschel Space
Observatory. By mapping the entire region, comparison with the
different Herschel-beams (9\arcsec--40\arcsec) will be
straight-forward.

The Serpens molecular core (also known as cluster A) is located at a
distance of 230$\pm$20 pc, following the discussion in
\citet{eiroa08}. The Serpens molecular core consists of several deeply
embedded sources, of which at least four are identified as containing
protostars \citep{wolf-chase98, hogerheijde99}, SMM1, SMM3, SMM4 and
S68N. Large-scale continuum-emission studies have been performed to
quantify the spectral energy distribution of all sources in order to
classify their evolutionary stage as well as the dust properties of
the molecular envelopes surrounding each source
\citep[e.g.,][]{casali93, hurt96, testi98, davis99, larsson00,
  williams00}, most recently with the Spitzer Space Telescope as part
of the Cores to Disks legacy program \citep[c2d;][]{harvey07,
  evans09}. These studies show that three of the sources (SMM3, SMM4
and S68N) have relatively low bolometric luminosities of
$\sim$~5~$L_\odot$ each, whereas SMM1 has a higher luminosity of
$\sim$~30~$L_\odot$ \citep[e.g.,][]{hogerheijde99, larsson00}. The
mass of each system (envelope and star) is in all cases less than
10~$M_\odot$. Recent interferometer observations by \citet{choi09}
show that SMM1 is a binary system with a projected separation of
$\sim$~500~AU. The binary SMM1b appears less embedded than the primary
(SMM1a). This discovery has been refuted by \citet{enoch09} and
\citet{vankempen09} who both resolve the disk around SMM1. Through
detailed SED modelling, \citet{enoch09} finds a very high disk mass of
$\sim$ 1 $M_\odot$ and that the inner parts of the envelope have been
cleared out to distances of 500 AU.

The region has been studied extensively at millimetre (mm) and
sub-millimetre (sub-mm) wavelengths in numerous molecular
transitions \citep[e.g.,][]{mcmullin94, white95, wolf-chase98,
  hogerheijde99, mcmullin00}, however it was not included in the
molecular surveys of isolated Class 0 and I objects in Perseus
and Ophiucus \citep{jorgensen04, jorgensen05, maret05}. The previous molecular
studies conclude that SMM1, SMM3, SMM4 and S68N are all very similar
to other young, low-mass stars with similar luminosities, such as
IRAS16293-2422 and NGC1333 IRAS4A and 4B in terms of abundances of
simple molecules that may be formed directly in the gas phase, e.g.,
HCO$^+$, CS, HCN \citep{mcmullin94, hogerheijde99,
  mcmullin00}. In Serpens, little has been done to quantify
excitation and gas-phase abundances of molecules predominantly formed
on dust grain surfaces, like CH$_3$OH, even though several lines have
been detected by \citet{mcmullin94, mcmullin00} and
\citet{hogerheijde99}.

More direct observations of grain surface products have been made by
\citet{pontoppidan04}, who mapped infrared absorption by molecules in
the ice over a region extending 40\arcsec\ south of SMM4, but still
located well within the molecular envelope. The primary ice
constituents were found to be H$_2$O, CO (0.4--0.9 with respect to
H$_2$O), CH$_3$OH (0.28 with respect to H$_2$O) and CO$_2$
\citep[0.3--0.5 with respect to H$_2$O;][]{pontoppidan08}. The
solid-state CH$_3$OH abundance is one of the
highest reported to date, both when compared to H$_2$O but also with
respect to gas-phase H$_2$ (3$\times$10$^{-5}$). At distances greater
than 12\,000~AU the CH$_3$OH-ice abundance drops beneath the detection
limit, corresponding to 3$\times$10$^{-6}$ with respect to H$_2$.

Besides the protostellar objects themselves, the region is permeated
by large-scale outflows extending several arcminutes from the
different sources with CO $J$=2--1 velocities ranging from $\pm$10--15
km\,s$^{-1}$ with respect to $\varv_{\rm LSR}=8-8.5$km\,s$^{-1}$
\citep[][Graves et al. subm.]{davis99}.
\citet{garay02} observed the outflows from SMM4 and
S68N in CH$_3$OH 3$_K$--2$_K$ emission and inferred CH$_3$OH
column densities of 1--2$\times$10$^{15}$~cm$^{-2}$, corresponding to
molecular abundance enhancements of $\sim$~50--330, depending on
outflow position, consistent with studies of other outflows
\citep[e.g.,][]{bachiller95, bachiller98}.

Here, the first map of rotationally excited methanol in the Serpens
Molecular Core is presented of transitions which cover the energy
range of $E_{\rm up} \sim 65-115$~K. The paper is structured as follows. In
Sect. \ref{sect:obs} the observations are presented, and in
Sect. \ref{sect:res} the observational results are
provided along with radiative transfer modelling. Section
\ref{sect:dis} presents a discussion of the results, with a particular
focus on the formation, desorption and excitation processes. Section
\ref{sect:conc} concludes the paper.

\section{Observations}\label{sect:obs}

\begin{figure*}
\includegraphics[width=\textwidth]{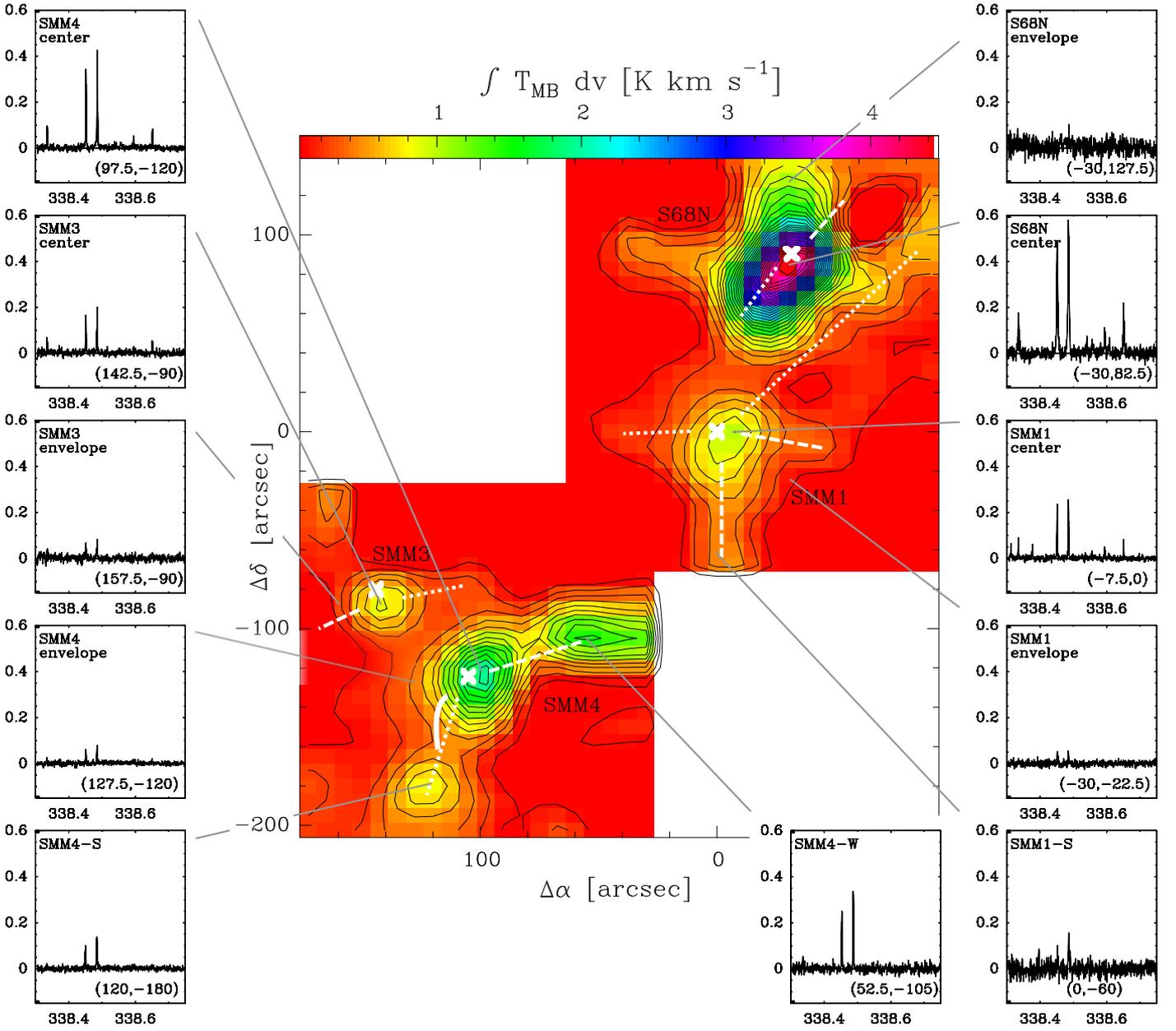}
\caption{Integrated emission, $\int T_{\rm MB}\, {\rm d}\varv$, of the
  strong 7$_0$--6$_0$ A$^+$ line at 338.41~GHz. Objects Serpens SMM1, 3, 4 and
  S68N are labelled. Offsets are in arcsec with respect to
  SMM1. Contours are at 3$\sigma$, 6$\sigma$, 9$\sigma$, etc. The
  units for colour are K\,km\,s$^{-1}$. The pixel size is
  7\farcs5$\times$7\farcs5. The lines indicate outflow directions,
  with dashed representing blue-shifted outflow emission, and dotted
  lines representing red-shifted emission. The full white line extending south
  of SMM4 depicts the location of the four stars used by
  \citet{pontoppidan04} to measure the CH$_3$OH-ice abundance. Shown
  are also sample spectra at different positions, illustrated by
  triangles, from a single spatial pixel and provided in each inset in
  units of arcsec. The abscissa is for frequency in units of
  GHz and the ordinate is for main beam temperature in units
  of K. All spectra have the same scale to ease intercomparison.}
\label{fig:find}
\end{figure*}

Observations of the Serpens molecular core were performed on June
20-22 2008 with the James Clerk Maxwell Telescope\footnote{The James
  Clerk Maxwell Telescope is operated by the Joint Astronomy Centre on
behalf of the Science and Technology Facilities Council of the United
Kingdom, the Netherlands Organisation for Scientific Research and the
National Research Council of Canada.} (JCMT) on Mauna Kea,
Hawaii. Observations were made of the 7$_K$-6$_K$ rotational band of
methanol ($E_{\rm up}$ $\sim$ 65--115~K) at frequencies ranging from
338 to 339 GHz using the Harp-B array receiver consisting of 4$\times$4
individual receivers \citep{smith03}. The telescope was pointed at two
different locations near Serpens SMM1 and two different locations near
SMM4, see Fig. \ref{fig:find}, covering a total extent of
4\farcm5$\times$5\farcm4. Observations were also made of C$^{18}$O,
$J$=3--2 at 329.330 GHz in a similar fashion, so as to be able to compare
methanol and CO emission. Observations were made in jiggle-mode to achieve
full spatial Nyquist-sampling using the Harp4 jiggle-pattern over a
2\arcmin$\times$2\arcmin\ region. The
beam-size of the JCMT at 338.5~GHz is $\sim$~15\arcsec\ and data were
subsequently put on a map with a pixel-size of 7\farcs5, i.e., one
half beam-size to fulfill the Nyquist sampling criterion. The
methanol observations were done using beam-switch with a throw of
180\arcsec\ in azimuth while the C$^{18}$O observations were done using position
switch to a clean position 1$^{\circ}$ off. The weather at the time of the
observations was good with
$\tau(\rm 225~GHz)\le 0.1$. Pointing and calibration were checked at
regular intervals, and the calibration error is estimated to be
$\lesssim$~20\% based on a comparison with standard calibration sources. Data were brought from the antenna temperature scale,
$T_A^*$, to the main beam temperature scale, $T_{\rm MB}$, by dividing
with the main beam efficiency, $\eta_{\rm MB}$, which is 0.60 for the
JCMT at these frequencies. The spectral resolution of the CH$_3$OH spectra is 0.43 km\,s$^{-1}$, and the C$^{18}$O spectra were rebinned to the same value.

The mean rms in the methanol spectra is 20~mK in 0.43~km\,s$^{-1}$
velocity bins over the
entire map.  The 1$\sigma$ noise level on integrated emission has been
determined as $1.2\,\sigma_{\rm rms}\,\sqrt{FWZI \times
  \delta\varv}$ where the factor 1.2 accounts for the 20\% telescope
calibration uncertainty, $\sigma_{\rm rms}$ is the rms noise, $FWZI$
the estimated full width at zero intensity (taken to be
5~km\,s$^{-1}$) and $\delta\varv$ the velocity bin
(0.43~km\,s$^{-1}$). The 3$\sigma$ uncertainty is typically $\sim$ 0.1
K\,km\,s$^{-1}$ per spatial pixel, but increases near the edges of the
map by a factor of 2--3.

Initial data reduction was done using the Starlink package. This
consisted of putting the spectra on to a regular grid with pixel-size
7\farcs5, subtracting linear baselines and co-adding all data-cubes. 
The Gildas package CLASS was used for subsequent analysis. 
Standard data reduction revealed emission in the off-position for a
subset of the methanol observations, but it
was possible to remove this emission due to two facts: (1) The
off-emission never coincided with on-emission, i.e., no
double-peaked or ``self-absorbed'' line profiles were observed. In
general the velocity offset between emission and absorption lines was
of the order of $\sim$~5~km\,s$^{-1}$ compared to typical line-widths
of 3--4~km\,s$^{-1}$ in the region where the off-emission was
seen. (2) Data were recorded with a shared off-position, that
is, even though the telescope jiggled on the on-position, no jiggling
was done on the off-position. Therefore it was possible to recreate
the off-emission in the strongest emission line at 338.409~GHz in the
following way: first the strongest ``absorption'' line was fitted with
a Gaussian profile, then an artificial spectrum was
created assuming a rotational temperature of 15~K, which was added to
each spectrum. The spectra were then analysed and no absorption
features remained within the noise limit.

\section{Results}\label{sect:res}

\begin{figure}
\includegraphics[width=\columnwidth]{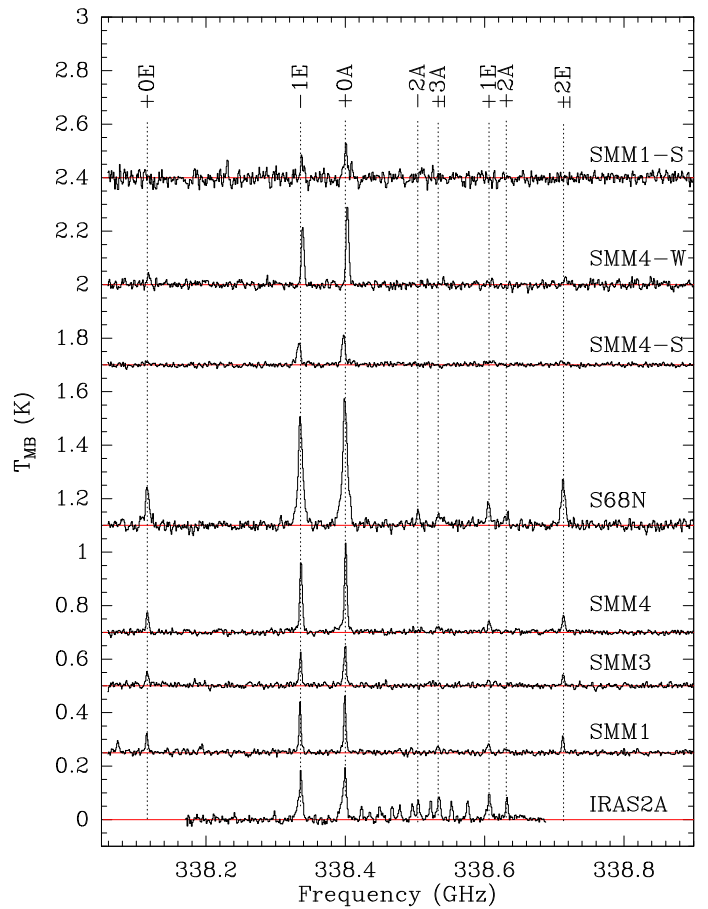}
\caption{Spectra of Serpens SMM1, 3, 4 and S68N. The NGC1333-IRAS2A
  spectrum is shown for comparison \citep{maret05}. Spectra at the
  three outflow positions, SMM1-S, SMM4-W and SMM4-S, are also shown. Lines
  identified in the CH$_3$OH 7$_K$--6$_K$ band are marked. Spectra
  were obtained by averaging emission from the peak pixel with its
  eight neighbouring pixels, i.e., in a box of
  22\farcs5$\times$22\farcs5. The SMM1, 3, 4 and S68N spectra are
  shifted by $+$0.25, 0.5, 0.7 and 1.1 K, respectively, while SMM1-S,
  SMM4-W and SMM4-S are shifted by 2.4, 2.0 and 1.7 K.}
\label{fig:spec}
\end{figure}

In all spectra emission lines from the 7$_K$--6$_K$
rotational band of methanol are detected both on and off sources. In
the central parts of each of the four YSOs up to eight emission lines
are detected originating from ten transitions. The strongest line, 
the 7$_0$--6$_0$ A$^+$ line at 338.409~GHz, has a peak brightness
temperature of up to 0.5~K. All line profiles are Gaussian, no line
asymmetries due to outflow activity or infall are observed. The $FWHM$
of the emission lines are $\sim$~4--6.5~km\,s$^{-1}$, consistent with
other observations of methanol line widths \citep[e.g.][]{maret05,
  jorgensen05}. Line widths change between objects, but remain
constant within each object, i.e., line widths do not change with
position within an envelope. In Fig. \ref{fig:find} an overview is
provided of the integrated emission in the strong 7$_0$--6$_0$ A$^+$
line over the Serpens molecular core. The integration is numeric and
has been done over the velocity interval --25 to $+$25 km\,s$^{-1}$
with respect to $\varv_{\rm lsr}=8.0$~km\,s$^{-1}$. The four Class 0
objects, Serpens SMM1, SMM3, SMM4 and S68N (labelled in
Fig. \ref{fig:find}) are clearly seen. Representative spectra obtained
at different positions sampling both YSOs, molecular envelopes and
outflows are shown. Besides the four Class 0 objects, three distinct
outflow knots are identified, SMM1-S, SMM4-W and SMM4-S. All of these
bright knots coincide with outflow positions as seen through CO
observations \citep[e.g.,][]{davis99}. Weaker structure is also seen
in the map, for example an elongation of SMM1 in the east-west
direction. This corresponds to the direction of a weaker CO-outflow
\citep{davis99}. The structure around S68N is more complex with the
source itself being elongated in the NW-SE direction. Around this
source a compact CS outflow has been discovered at the same position
angle as the elongation \citep{wolf-chase98}. Weak emission is also
seen around S68N in the north-south and east-west directions.

Representative spectra of the
central parts of the four envelopes are shown in Fig.~\ref{fig:spec}
where emission is averaged over 3$\times$3 pixels (corresponding to 
22\farcs5$\times$22\farcs5 or 5600~AU$\times$5600~AU). In the
following, results will be presented first for the central parts of
each envelope followed by results for the extended emission in each
envelope. Finally results will be presented for three strong outflow
knots in the map, named SMM1-S, SMM4-W and SMM4-S.

\begin{table*}
\caption{Integrated line intensities, $\int
  T_{\rm MB}\ {\rm d}\varv$ (K\,km\,s$^{-1}$) over the central
  22\farcs5 $\times$22\farcs5 of each source$^a$ and
  emission-weighted, average line-widths.}
\label{tab:line}
\center\begin{tabular}{l r r r r r r | r r r}
\hline\hline
Transition & $\nu$ (GHz)$^b$ & $E_{\rm up}$ (K)$^b$ & SMM1 & SMM3 &
SMM4 & S68N & SMM1-S & SMM4-W & SMM4-S \\
\hline
+0E     & 338.1245 &  78.1 & 0.25 & 0.22 & 0.30 & 0.85 & \ldots & 0.09 & 0.41 \\
-1E     & 338.3446 &  70.5 & 0.58 & 0.45 & 1.09 & 2.90 & 0.52 & 0.90 & 1.38 \\
+0A     & 338.4087 &  65.0 & 0.71 & 0.62 & 1.35 & 3.66 & 1.06 & 1.26 & 2.21 \\
-2A     & 338.5129 & 102.7 & 0.06 & 0.01 &  \ldots & 0.24 & 0.08 & 0.20 & \ldots \\
$\pm$3A$^c$ & 338.5420 & 114.8 & 0.11 & 0.16 & 0.15 & 0.40 & \ldots & \ldots & \ldots \\
+1E     & 338.6150 &  86.1 & 0.16 & 0.08 & 0.18 & 0.45 & \ldots & 0.34 & \ldots \\
+2A     & 338.6399 & 102.7 & 0.09 & 0.07 &  \ldots & 0.26 & \ldots & \ldots & \ldots \\
$\pm$2E$^c$ & 338.7223 &  89.1 & 0.21 & 0.16 & 0.28 & 1.05 & \ldots & 0.13 & 0.11 \\
\hline
$\Delta\varv$ (km\,s$^{-1}$) & & & 3.4$\pm$0.9 & 4.4$\pm$2.5 & 4.0$\pm$0.7 & 7.2$\pm$1.3 & 4.3$\pm$0.9 & 4.0$\pm$0.3 & 4.1$\pm$0.3 \\
\hline
\end{tabular}
\flushleft{$^a$ Non-detections are marked by ``\ldots'', for which an
  upper limit of 0.03~K\,km\,s$^{-1}$ applies.\\
$^b$ Frequencies and upper level energies are from the
  Cologne Database for Molecular Spectroscopy \citep[CDMS;][]{muller01}.}
\\$^c$ Lines are blended. The frequencies and upper level energies are
average values.
\end{table*}

\subsection{Emission from the central part of the envelopes}

\begin{figure}
\includegraphics[width=\columnwidth]{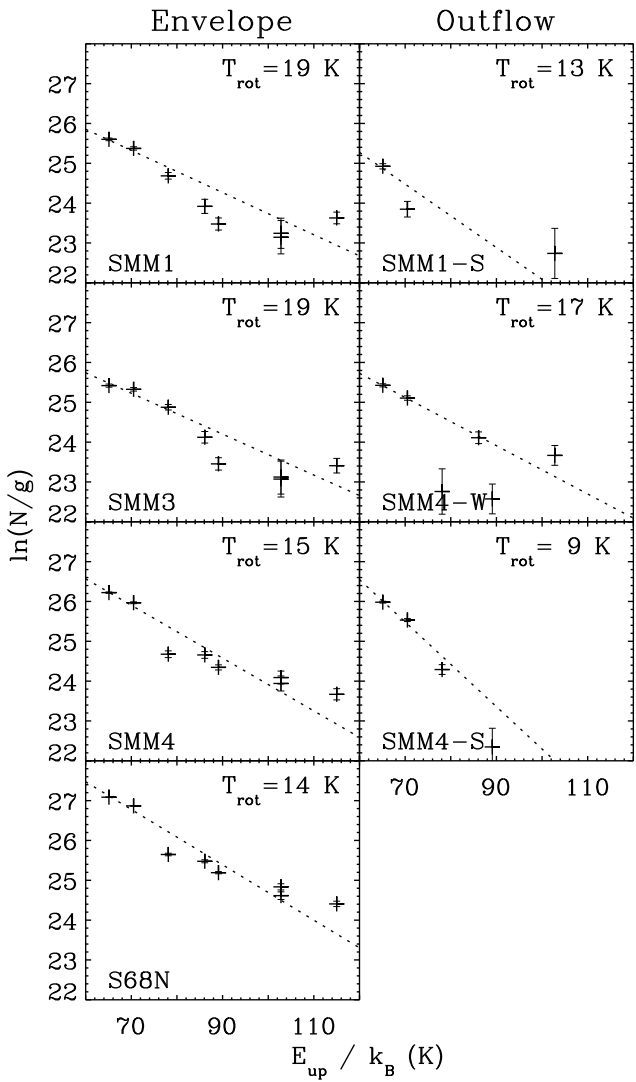}
\caption{Rotational diagrams with data obtained from the four spectra
  shown in Fig. \ref{fig:spec}. The best-fit straight line is shown in
each case as a dotted line. Error bars are for 1$\sigma$
uncertainties.}
\label{fig:rot}
\end{figure}

The emission lines arising from the central part of each envelope are
fitted with Gaussian line profiles to obtain the integrated emission
at higher accuracy. Results are tabulated in Table~\ref{tab:line}
along with the upper level energy, $E_{\rm up}$, and line
frequency. In the spectra presented in Fig. \ref{fig:spec} the
3$\sigma$ level is a factor of three lower, since they are each the
average of nine spectra. Low-$K$ lines up to $K$ = 3 are clearly
detected, but high-$K$ lines are not seen. This is in contrast
to other low-mass YSOs with similar luminosities and/or distances and evolutionary stages, such
as IRAS16293-2422 \citep[e.g.][]{vandishoeck95} and NGC1333 IRAS2A
\citep{maret05}.

The integrated intensities are used to make rotational diagrams for
each of the four objects. These are shown in
Fig. \ref{fig:rot}. Within the uncertainty, the logarithm of the
derived upper
level column densities, $N$, divided by the statistical weights, $g$,
fall on a straight line when plotted versus the upper level
energy divided by the Boltzmann constant, $k_{\rm B}$. No difference
between A- and E-type methanol is found, thus
the abundances are equal, \mbox{$x$(E-CH$_3$OH) = 
$x$(A-CH$_3$OH)}. In the cold-temperature limit ($T \sim$ 10 K) the abundance ratio is expected to be $E/A=0.69$ whereas it is 1 at higher temperatures, and thus little or no difference in abundance is expected \citep[e.g.,][]{friberg88}. The derived rotational temperature for each of the
four envelopes is low, 15--20~K, significantly
lower than the rotational temperature of $\sim$ 80~K inferred for
IRAS16293-2422 and NGC1333 IRAS2A \citep{vandishoeck95, maret05}. The
estimated column densities
are all of the order of $\sim$~10$^{14}$--10$^{15}$~cm$^{-2}$. Results
are tabulated in Table \ref{tab:rot}. Rotational temperatures and
column densities are similar to those found by \citet{maret05} and
\citet{jorgensen05} for a larger sample of low-mass, Class 0 objects,
with the exception of the two sources mentioned above.

\subsection{Extended emission from the envelopes}

To quantify the spatial distribution of methanol in each molecular
envelope, the radial distribution of emission from the five
strongest lines is plotted in Fig. \ref{fig:column}. Methanol emission is
extended in all envelopes, and the $FWHM$ of emission is around
2--3 times the 15\arcsec\ beam. Extended methanol has previously been
reported in the envelope of the isolated Class 0 object L483
\citep{tafalla00} in the 2$_0$--1$_0$ A$^+$ and 2$_{-1}$--1$_{-1}$ E
lines ($E_{\rm up}$~=~12--20~K), where it was found that the methanol
emission traces total gas column density. To verify whether this is
true for the Serpens  envelopes, it is necessary to quantify the
physical structure of the envelopes. This is done through a
combination of dust continuum emission at 850 $\mu$m as observed with
SCUBA on the JCMT and modelling of the physical structure.

To obtain the absolute abundances, three different methods are
used. First, the CH$_3$OH column density obtained from rotational
diagrams is compared to the column density obtained from dust emission
at 850 $\mu$m in the same beam. Second, the CH$_3$OH column density
from rotational diagrams is compared to the column density predicted
from a physical model of each source. Third, a radiative transfer
model is compared directly to the observed spectrum to constrain the
abundance. Results are listed in Table \ref{tab:abun}.

\begin{table}
\caption{Rotational diagram results for the observed methanol emission.}
\label{tab:rot}
\center\begin{tabular}{l r r}
\hline\hline
Source & $T_{\rm rot}$ (K) & $N$ (10$^{14}$~cm$^{-2}$)$^a$ \\
\hline
SMM1 & 18.9$\pm$0.9 & 4.1$\pm$0.7 \\
SMM3 & 19.5$\pm$1.1 & 3.4$\pm$0.7 \\
SMM4 & 15.1$\pm$0.4 & 12$\pm$1.4 \\
S68N & 14.4$\pm$0.2 & 34$\pm$1.7 \\
\hline
SMM1-S & 12.6$\pm$2.4 & 5.7$^{+5.8}_{-5.7}$ \\
SMM4-W & 16.5$\pm$1.3 & 4.4$\pm$1.4 \\
SMM4-S & 9.3$\pm$1.0 & 34$\pm$27 \\
\hline
\end{tabular}
\flushleft{$^a$ Column density is the sum of A- and E-type methanol
  column densities.}
\end{table}

\subsubsection{Dust continuum emission}\label{sec:dust}

It is possible to directly estimate the average column density over
the central area of each envelope by assuming that the dust emission
is optically thin, thermal emission at a single dust temperature,
$T_{\rm d}$:
\begin{eqnarray}
\bar{N} &=& \frac{S_\nu d^2}{\kappa_\nu B_\nu(T_{\rm d})A} \nonumber\\
&=& 4.90 \times 10^{16} {\rm \ cm}^{-2}\ S_\nu({\rm Jy})\, d({\rm pc})^2
\left[ \exp \left(\frac{16.9~{\rm K}}{T_{\rm d}} \right) - 1 \right]\ .
\end{eqnarray}
Here $S_\nu$ is the continuum flux integrated over
22\farcs5$\times$22\farcs5 at 850~$\mu$m, $d$ the distance to the
object, $\kappa_\nu$ the dust opacity at 850~$\mu$m taken from
\citet{ossenkopf94}, $B_\nu$ the Planck function at a dust temperature
$T_{\rm d}$ and $A$ the total surface area, in this case
7.1$\times$10$^{33}$~cm$^2$. Finally a gas/dust ratio of 100 is
assumed.

Reduced data from the SCUBA Legacy archive have been used
\citep{difrancesco08} to estimate the integrated continuum flux at
850~$\mu$m. An average dust temperature of 20~K was used following
global estimates of the dust temperature over the entire Serpens
molecular core of 20~K from \citet{schnee05}. If the temperature
changes to 10~K, the column density estimate is increased by a factor
of 3.3, while if it is 30~K, the column density is decreased by a
factor of 1.8.

By using a dust temperature of 20~K, the average column density over a
22\farcs5 $\times$ 22\farcs5 region is in the range of 1--4 $\times$
10$^{23}$~cm$^{-2}$. This leads to CH$_3$OH fractional abundances in
the range of 1--20 $\times$ 10$^{-9}$ with respect to H$_2$, see
below.

\subsubsection{{\sc Dusty} modelling of envelope properties}

Since the dust (and gas) temperatures change near the YSOs, the
CH$_3$OH emission can be further quantified through modelling of the
physical parameters of the envelopes. This is done following
\citet{schoier02} and \citet{jorgensen02}, where the dust
continuum emission is modelled with a spherically symmetric envelope
with a power-law density structure heated from the inside by the
protostar, using the 1D code {\sc Dusty} \citep{ivezic97}. Dust opacities
tabulated by \citet{ossenkopf94} are used for densities of
10$^6$~cm$^{-3}$ and thin ice mantles (corresponding to the values in
their Table 1, column 5). The central heating
source is taken to be a black-body radiating at a temperature of
5000~K, but model results are not very sensitive to this
parameter. The output from {\sc Dusty} is dimensionless, so to make an
absolute calibration it is necessary to do so against the absolute
luminosity and a given distance of the source. The
output can be in the form of a spectral energy distribution (SED) and
a radial profile of continuum emission at a user-specified wavelength. 

Only dust heating at temperatures below 250~K is considered
(corresponding to a peak wavelength of $\sim$~12~$\mu$m). Thus there
is a ``hole'' in the inner part of the envelope extending out to a
radius, $r_{\rm in}$, where no attempt is made to model the dust
emission. The physical extent of the envelope is defined
by the parameter $Y=r_{\rm in}/r_{\rm out}$. The density
profile of the envelope is described by a power-law such that $n(r)
\propto r^{-p}$. Three
observational constraints are used following \citet{jorgensen02} to
identify the best-fitting dust model: the SED and the spatial
distribution of continuum emission as recorded by SCUBA on the JCMT at
450~$\mu$m and 850~$\mu$m, both available through the SCUBA legacy
archive \citep{difrancesco08}.

SEDs were assembled from the literature, using data
points from $MIPS$ on the Spitzer Space Telescope, ISO-LWS and IRAS
along with ground-based single-dish continuum measurements at
sub-mm, mm and cm wavelengths. For this work, the SED database
available at \url{http://astro.kent.ac.uk/protostars/} proved
invaluable \citep{froebrich05}. A full table of SED points is
included in Appendix A.

The reduced SCUBA continuum emission maps from the SCUBA legacy
archive have been smoothed to equivalent beam sizes of 11\arcsec\ and
19\farcs5 at 450 and 850~$\mu$m, respectively
\citep{difrancesco08}. Along with emission maps, maps containing the
error on each pixel are also provided, facilitating error
analysis. The radial profile of each object was compiled by
considering the emission in annuli extending from the pixel
containing the maximum value of emission. The maps were carefully
checked, and directions containing contamination from other objects
were avoided.

\begin{figure}
\includegraphics[width=\columnwidth]{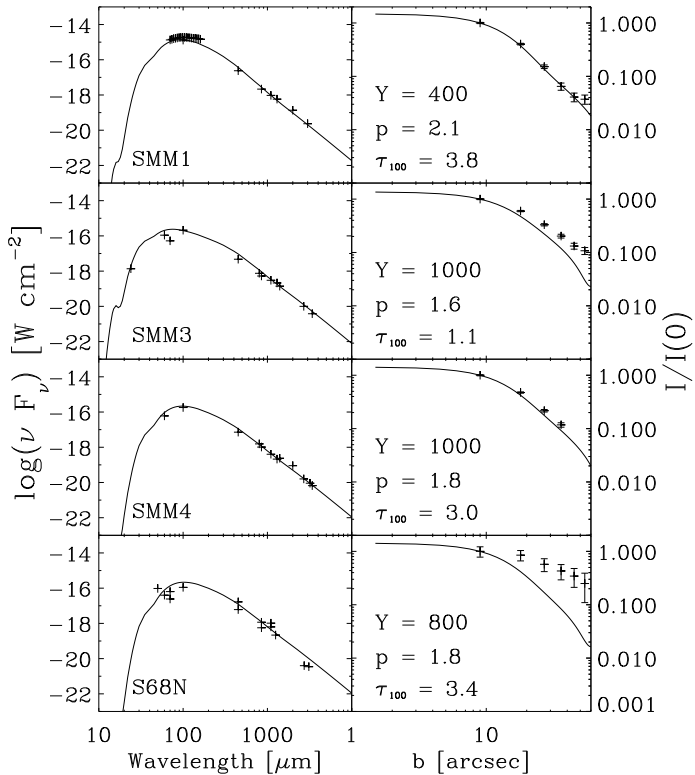}
\caption{{\sc Dusty} modelling results of the molecular envelopes of SMM1,
  3, 4 and S68N. The left panel shows the SEDs of all four objects,
  whereas the right panel shows the normalized radial dust emission
  profile as obtained with SCUBA \citep{difrancesco08}. Best-fit
  results are over-plotted as solid lines, and the values for the
  best-fit models are given in the right-hand panel.}
\label{fig:dusty}
\end{figure}

Three {\sc Dusty} input values were varied: the physical size of the
envelope, $Y$, the opacity at 100~$\mu$m, $\tau(100~\mu$m) and the
power-law slope of the radial density profile, $p$. A small grid of
simulations was calculated and the best-fit model for each envelope
was found using a $\chi^2$ method, following \citet{jorgensen02}. The
best-fit model parameters are listed in Table \ref{tab:dusty} along
with physical parameters such as the extent of the envelope, the local
density at the inner radius, the column density and the total envelope
mass. The best-fit model predictions are overplotted on actual data in Fig. \ref{fig:dusty}.

In general agreement between the modelled and observed properties are
good, except in the case of the modelled 850 $\mu$m radial profile of
S68N. The model prediction is consistently a factor of 2--3 lower than
observed values. Part of the reason is that since several sub-mm
sources are present close to S68N, only emission from the NW direction
is modelled here, implying that the uncertainty on individual data
points is higher than in the case of SMM1, SMM3 and SMM4. In this
particular direction there may still be contamination, and thus
emphasis is placed primarily on modelling of the SED. Moreover, the
radial emission profile at 450 $\mu$m is much steeper than the 850
$\mu$m profile, indicating that there may be more cold, ambient cloud 
material toward S68N. The radial profile obtained at 450 $\mu$m is
well reproduced by the best-fit {\sc Dusty} model.

\begin{table*}
\caption{Best fit {\sc Dusty} model input parameters and results for the
  physical structure of the four molecular envelopes in Serpens.}
\label{tab:dusty}
\center\begin{tabular}{l r r r r | r}
\hline\hline
Property & SMM1 & SMM3 & SMM4 & S68N & IRAS2A \\
\hline
$L$ ($L_\odot$) & 37 & 2.0 & 6.3 & 3.6 & 21 \\
$d$ (pc) & \multicolumn{4}{c|}{230} & 250\\
\hline
$p$ & 2.1 & 1.6 & 1.8 & 1.8 & 1.9\\
$\tau$(100~$\mu$m) & 3.8 & 1.1 & 3.0 & 3.4 & 1.8\\
$Y$ (=$r_{\rm in}$/$r_{\rm out}$) & 400 & 1000 & 1000 & 800 & 500 \\
\hline
$r_{\rm in}$ (AU) & 49 & 14 & 17 & 18 & 31.3 \\
$r_{\rm out}$ (10$^4$~AU) & 2.0 & 1.4 & 1.7 & 1.4 & 1.6 \\
$n_{\rm in}$ (10$^9$~cm$^{-3}$) & 1.7 & 0.94 & 2.8 & 3.0 & 1.0 \\
$n_{\rm 1000~AU}$ (10$^6$~cm$^{-3}$) & 3.0 & 1.0 & 1.9 & 2.2 & 1.4 \\
$N$(H$_2$) (10$^{24}$~cm$^{-2}$) & 1.1 & 0.33 & 0.89 & 1.0 & 0.53 \\
Envelope mass, $T >$ 10~K ($M_\odot$) & 4.1 & 2.5 & 3.9 & 3.6 & 2.2 \\
\hline
\end{tabular}
\end{table*}

From the {\sc Dusty} model it is possible to estimate the average column
density over a region of 22\farcs5 $\times$ 22\farcs5 and compare this
to the CH$_3$OH column density. The average column density is
found to be $\sim$ (1.0 $\pm$ 0.2) $\times$ 10$^{23}$
cm$^{-2}$. This leads to an abundance of $\sim$ 3 $\times$
10$^{-9}$ with respect to H$_2$ for SMM1 and SMM3 to $\sim$ 30
$\times$ 10$^{-9}$ for S68N (see Table \ref{tab:abun} for
details).

\subsubsection{{\sc Ratran} simulation}

In the following the abundance is constrained through direct radiative
transfer modelling. This has been done using the code, {\sc Ratran}
\citep{hogerheijde00} in conjunction with
molecular data from LAMDA \citep{schoier05}. This opportunity was used
to update collisional data in the LAMDA database for
methanol using newly released rate coefficients for collisions between
para-H$_2$ and A- and E-type
methanol up to and including $J=9$, corresponding to an upper level
energy of $\sim$~500~K \citep{pottage04}. The rate coefficients are
given for temperatures from 5 to 200~K. The
coefficients are not complete up to upper level energies of 500~K
since they are
limited to $J<$~10. In the case of A-type methanol with $K=0$, this
corresponds to an upper level energy of $\sim$ 100~K. Data have been
extrapolated so that they are complete up to an upper level
energy of 385~K corresponding to the first torsional state. For the
extrapolation only transitions between $K$-ladders
and $\Delta J=1-5$ are considered. The collisional rate coefficients
are found to be proportional
to $\sqrt{T}/\Delta J$ which is used for the extrapolation
\citep{leurini07}. Level energies and Einstein $A$-values are taken
from the Cologne Database for Molecular Spectroscopy
\citep[CDMS;][]{muller01}.

The density and temperature profile of the {\sc Dusty} modelling is
used for the physical structure of the envelope. No effort is made to
model the line widths, but instead a constant, turbulent line width of
3.0--5.0~km\,s$^{-1}$ is assumed, depending on source. This is not a
measure of the actual turbulence in the region, but only a means to
ensure that the modelled linewidth corresponds to the observed. For
each source a small grid of {\sc Ratran} models has been run with a
so-called ``jump'' abundance structure. The jump was located at
$T_{\rm dust} = $80~K ($R$ = 50--125 AU) with an increase in the
abundance of a factor of 1--10$^{3}$ corresponding to most CH$_3$OH
evaporating off of the grains. When running the {\sc Ratran} models,
great care has been taken to ensure that the central region containing
the jump is properly sampled. In order to do this the pixel-size is
set to 0\farcs2 (45~AU) and the central pixels were oversampled by a
factor of 50.

\begin{figure*}
\sidecaption
\includegraphics[width=12cm]{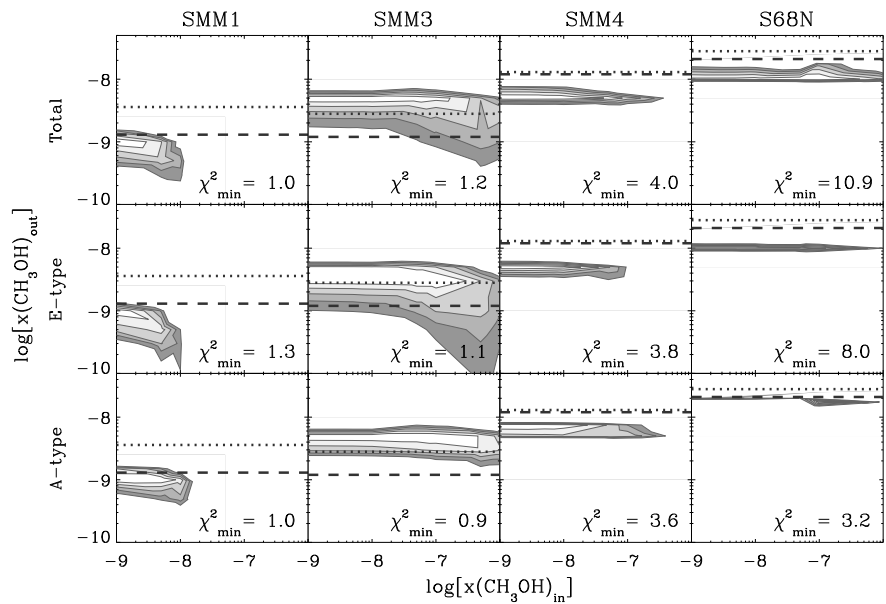}
\caption{Reduced $\chi^2$ distribution as a function of inner and
  outer abundance for A- and E-type CH$_3$OH as well as the combined
  value. Contours are for 1, 2, 3, 4 and 5$\sigma$. The dotted line
  corresponds to the abundance derived using $N$(H$_2$) from {\sc
    Dusty} modelling, and the dashed line is the abundance derived
  using $N$(H$_2$) directly from SCUBA dust continuum emission. The
  minimum $\chi^2_{\rm red}$ is given in each panel.}
\label{fig:chi2}
\end{figure*}

A $\chi^2$ method is then used to determine the best-fit model
integrated emission compared to observed integrated emission. This is
done separately for A- and E-type CH$_3$OH to examine relative
abundances. No significant difference is found  between A- and E-type
CH$_3$OH abundances, thus confirming the results obtained from the
rotational diagrams. All abundances are listed in Table
\ref{tab:abun}. Diagrams illustrating the variation of $\chi^2$ with
inner and outer abundance are shown in Fig. \ref{fig:chi2}. Here it
may be seen that the inner abundance is not well constrained, and it
is only possible to provide upper limits ranging from
2$\times$10$^{-9}$ (SMM1) to
5$\times$10$^{-7}$ (SMM3 and S68N). On the other hand, the outer
envelope abundance is very well constrained for all sources and lies
in the range of 10$^{-9}$ (SMM1) to greater than 10$^{-8}$ (S68N). The
methanol enhancements, $x_{\rm in} / x_{\rm out}$, are from $\lesssim$
2--3 (SMM1) to $\lesssim$ 200 (SMM3). Due to the lack of high-$K$
lines it is not possible to further constrain the inner abundance.

\begin{table*}
\caption{Average total gas column densities and fractional abundances
  of CH$_3$OH.}
\label{tab:abun}
\center\begin{tabular}{l r r r r r r}
\hline\hline
Source & $N_{\rm DUSTY}^a$ & $N_{\rm SCUBA}^b$ &
$x$(CH$_3$OH)$_{\rm DUSTY}^c$ & $x$(CH$_3$OH)$_{\rm SCUBA}^c$ &
$x$(CH$_3$OH)$_{\rm out}^{c,d}$ & $x$(CH$_3$OH)$_{\rm in}^{c,d}$\\
 & (10$^{23}$~cm$^{-2})$ & (10$^{23}$~cm$^{-2}$) & $\times10^{-9}$ &
$\times10^{-9}$ & $\times10^{-9}$ & $\times10^{-8}$ \\
\hline
SMM1 & 1.3  &   3.5$\pm$0.3 & 3.6$\pm$0.9 & 1.3$\pm$0.4 & 1 & $\le$0.3 \\
SMM3 & 0.83 & 2.0$\pm$0.13 & 2.8$\pm$0.7 & 1.2$\pm$0.3 & 4 & $\le$50  \\
SMM4 & 1.1  & 1.2$\pm$0.07 & 13$\pm$3    & 12$\pm$3    & 5 & $\le$5  \\
S68N & 1.2  & 1.7$\pm$0.10 & 28$\pm$2    & 21$\pm$2    & 10 & $\le$40 \\
\hline
\end{tabular}
\flushleft{$^a$ Gas column density, $N$(H$_2$), obtained from {\sc Dusty}
  modelling over a region of 22\farcs5$\times$22\farcs5 assuming a
  gas:dust ratio of 100.\\
$^{b}$ Gas column density , $N$(H$_2$),obtained from SCUBA emission at
  850 $\mu$m
  assuming a constant dust temperature of 20~K.\\
$^c$ Fractional CH$_3$OH abundance summed over A- and E-type CH$_3$OH.\\
$^d$ Inner and outer abundance in the {\sc Ratran} jump-model.
}
\end{table*}

\subsubsection{Comparison of abundance measurements}

The three methods for determining the abundance may be divided into
the following categories: 
\begin{enumerate}
\item Direct observational abundance assuming LTE in terms of CH$_3$OH
  excitation and optically thin dust emission.
\item Combined direct and model abundance, still assuming LTE in terms
  of CH$_3$OH excitation but modelling the dust emission with a range
  of temperatures.
\item Modelling of the CH$_3$OH emission within a physical, spherically
  symmetric model with varying dust and gas temperatures and densities.
\end{enumerate}
Despite the obvious differences and levels of complexity between the
three methods, the results in terms of average methanol abundances are
remarkably similar. Typically, variations are within a factor of 2--3.

In the following only results from the {\sc Ratran} modelling will be
discussed. Since this method takes the full density and temperature
structure into account it will be the more accurate of the three as
demonstrated for other molecules \citep[e.g.,][]{hogerheijde00,
  jorgensen02, jorgensen04}.

\subsection{Outer envelope abundance structure}\label{sect:outenv}

\begin{figure*}
\includegraphics[width=\textwidth]{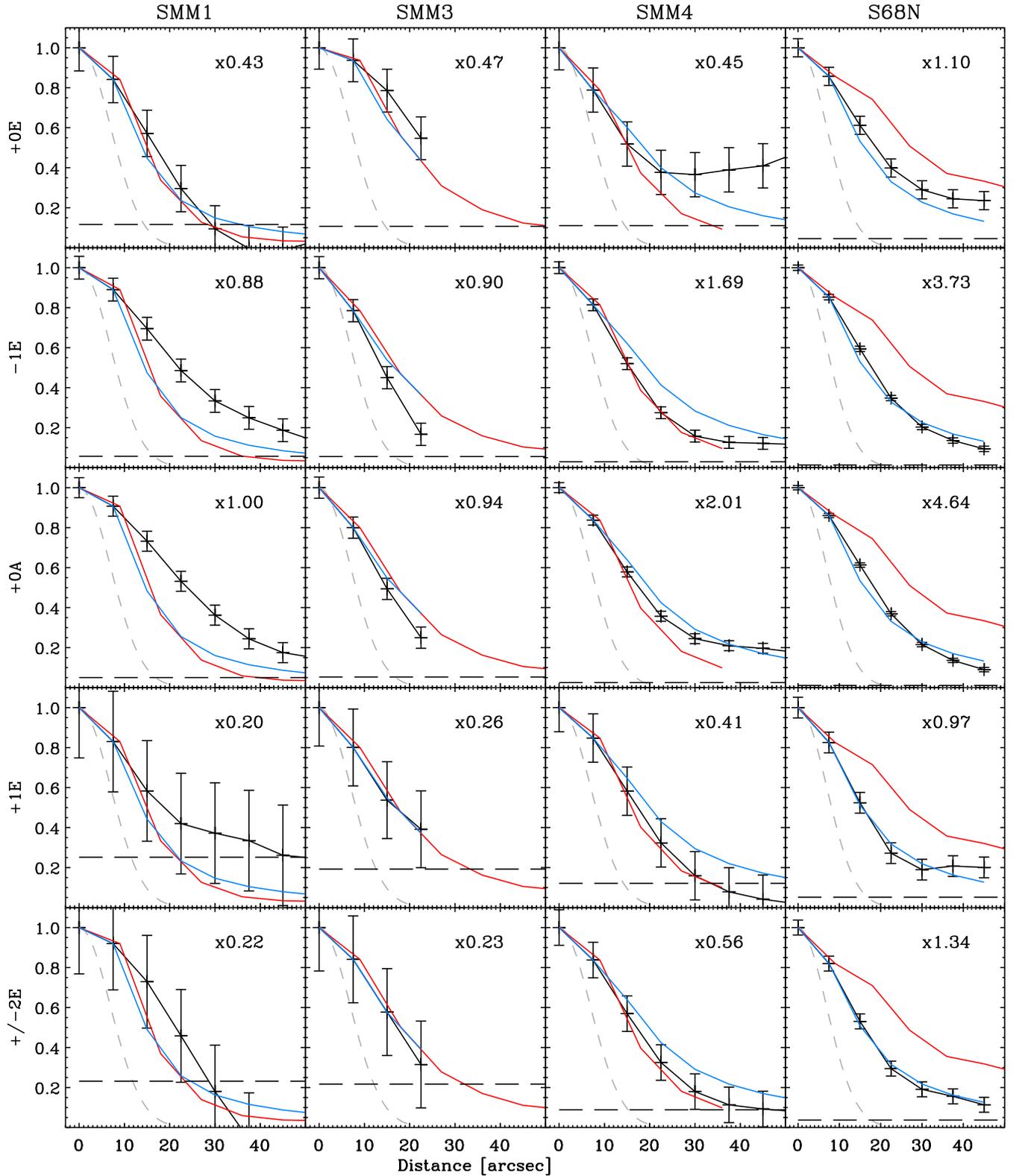}
\caption{Spatial distribution of methanol emission in the four
  objects, Serpens SMM1, 3, 4 and S68N. The integrated emission of the
  five lines 7--6 +0E, --1E, +0A, +1E and $\pm$2E is shown
  (black). The error-bars are for 1$\sigma$ uncertainties. Overlaid is
  the dust continuum emission as obtained from
  SCUBA (red) and the {\sc Dusty} dust continuum modeling column density
  (blue). All profiles have been normalized to 1. The normalization
  factor for the peak emission is shown in each diagram. The
  3$\sigma$ noise level is shown as a dashed line, and the grey, dashed line
  shows the beam profile.}
\label{fig:column}
\end{figure*}

In Fig.~\ref{fig:column} the column densities as predicted by
the {\sc Dusty} modelling and the 850~$\mu$m dust continuum emission radial
profiles are overlaid. The methanol emission profiles follow both the
dust emission and column density profiles very closely over scales
from $\sim$ 5500 AU to greater than 12\,000 AU. Because the
distribution is independent of rotational line, the rotational
temperature (corresponding to line ratio) is constant throughout the
envelope and the methanol column density scales directly with the
envelope column density, similar to the case of L483
\citep{tafalla00}. Thus, over this range of radial distances, the
average CH$_3$OH gas abundance is constant. In the case of S68N the
agreement between the observed and simulated radial profiles is not
good since the simulation under-estimates the emission at larger
radii. Furthermore, the outer envelope abundance as derived from
rotational diagrams or from comparison with dust emission agrees to
within a factor of 2 with the outer abundance as derived from the {\sc
  Ratran} modelling.

\subsection{Emission from outflow positions}

\begin{figure}
\center\includegraphics[width=0.9\columnwidth]{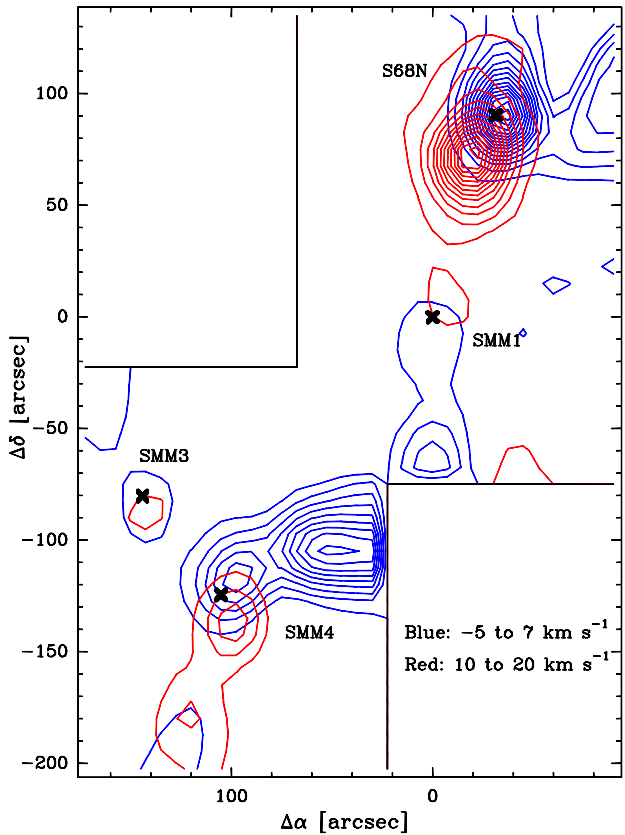}
\caption{Outflow emission in Serpens as traced by the CH$_3$OH
  7$_0$--6$_0$ A$^+$ line. Contours are for red- and
  blue-shifted outflow emission and are at 3$\sigma$, 6$\sigma$,
  9$\sigma$, \ldots\ Crosses mark the positions of SMM1, SMM3, SMM4 and
  S68N.}
\label{fig:ch3oh_out}
\end{figure}

\begin{figure}
\center\includegraphics[width=\columnwidth]{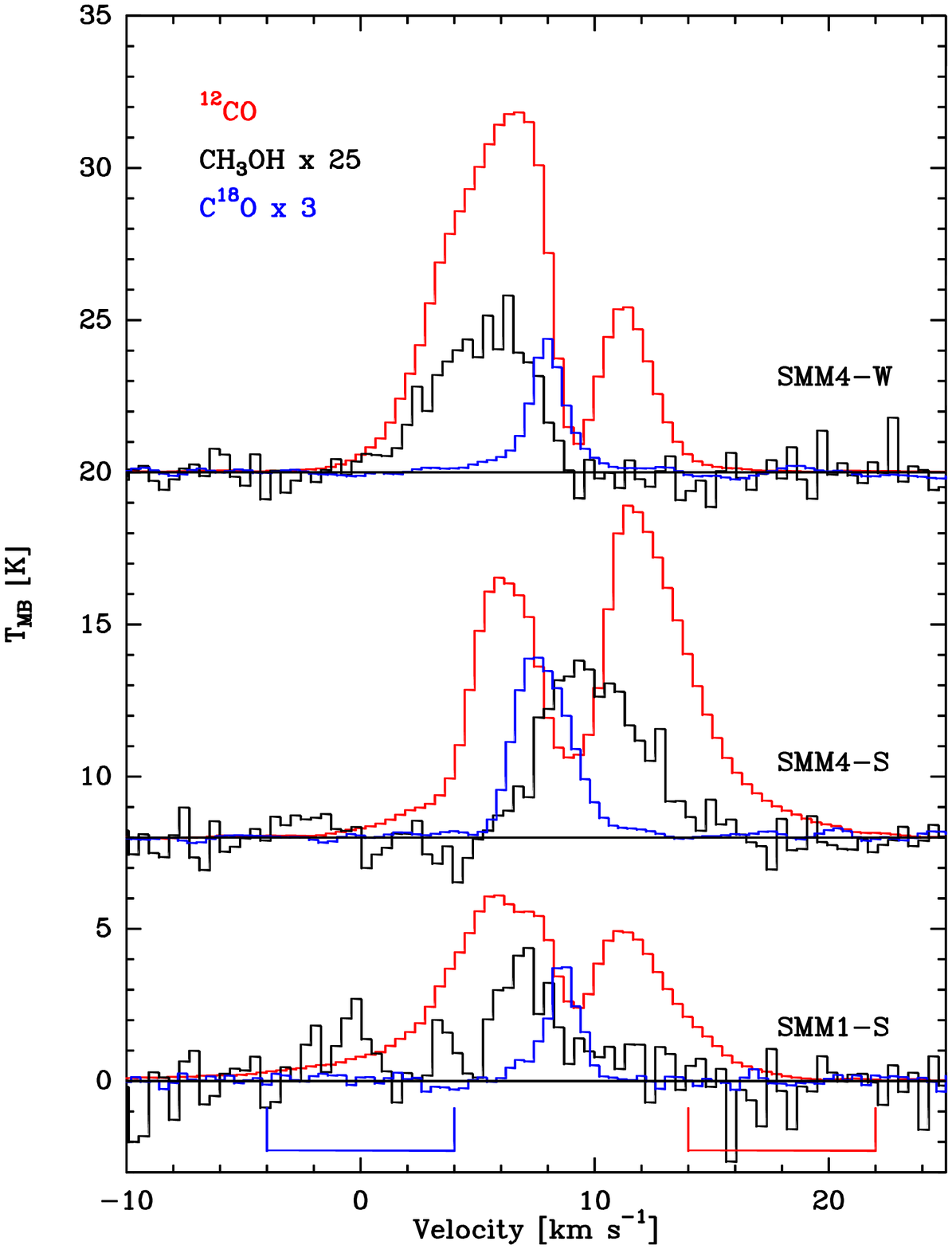}
\caption{Comparison of outflow line profiles. For each outflow
  position, SMM1-S, SMM4-W and SMM4-S, line profiles of $^{12}$CO
  $J$=3--2 (red), C$^{18}$O $J$=3--2 (blue) and CH$_3$OH
  7$_0$--6$_0$ A$^+$ (black) are shown. C$^{18}$O and CH$_3$OH spectra
  have been multiplied by 3 and 25, respectively, for clarity. SMM4-S
  and SMM4-W spectra have been shifted by 8 and 20 K. The masks at the
 bottom indicate the velocity range over which $^{12}$CO emission is
          integrated to obtain the outflow column density.}
\label{fig:co_out}
\end{figure}

Methanol is detected along the outflows throughout the emission map
(Fig. \ref{fig:find}). In the following, focus will be placed on
emission from three distinct knots labelled SMM1-S, SMM4-W and SMM4-S
on Fig. \ref{fig:find}. The elongated emission associated with
S68N indicates
that a compact shock is responsible for sputtering CH$_3$OH from the
grain mantles into the gas phase. This outflow has already been
observed in CS $J$=3--2 emission \citep{wolf-chase98}. The line
profiles associated with this outflow are symmetric and Gaussian in
nature making an accurate disentanglement of the contributions from
envelope and outflow difficult. Unlike the above mentioned outflow
knots, emission is spatially coincident with the source position. For
this reason the S68N outflow is not included in the following
discussion.

To investigate the properties of each outflow knot, spectra are
extracted from the map and rotational diagrams are made in the same
manner as above (see Figures \ref{fig:spec} and \ref{fig:rot}). 
The intensity of the
strong 7$_0$--6$_0$ A$^+$ line is comparable to the emission from SMM4
in all three knots. The detected transitions and associated intensities are
listed in Table \ref{tab:line}. The lines are shifted  by $\sim$
1~km\,s$^{-1}$ with respect to each other and with respect to the
source velocities. Line profiles are symmetric and Gaussian in nature,
with no obvious line-wings. Nevertheless, methanol clearly traces the
outflow activity associated with SMM1, SMM4 and S68N as seen in
Fig. \ref{fig:ch3oh_out}, whereas little emission is associated with
SMM3. In Fig. \ref{fig:ch3oh_out} the contributions from red- and
blue-shifted emission are estimated by integrating over the velocity
intervals from --5 to 7~km\,s$^{-1}$ and 10 to 20~km\,s$^{-1}$,
respectively. This is to be compared with the average source velocity,
$\varv_{\rm lsr}=8.5$~km\,s$^{-1}$.

Results obtained from the rotational diagrams are tabulated in Table
\ref{tab:rot}. The rotational temperature is $\sim$ 10--15 K,
indicating sub-thermal excitation at all three outflow positions,
as found typically also in other outflows because of the relatively
low density compared with the central envelope
\citep[e.g.,][]{bachiller95}. The column densities range
from a few $\times$~10$^{14}$~cm$^{-2}$ (SMM1-S and SMM4-W) to
$\sim$ 3$\times$10$^{15}$~cm$^{-2}$ (SMM4-S). To obtain accurate
abundance estimates of methanol in shocked regions, it is again
imperative to estimate the gas column density accurately. In this case 
$^{12}$CO $J$=3--2 data from the JCMT science archive, obtained in the
context of the JCMT Gould Belt legacy survey, are used (Graves et
al. subm.). The principle for obtaining the column density from
$^{12}$CO in the shocked gas is that emission from the higher
velocity line-wings is optically thin. This assumption has been shown
to be valid for a number of molecular outflows \citep[see e.g.,][ for
  a review]{bachiller99}. In Fig. \ref{fig:co_out} the line profiles
are shown for the three outflow positions analysed here. Emission from
SMM1-S and SMM4-W is blue-shifted, while emission from SMM4-S is
red-shifted. The CO line-wings are not as pronounced as in other
well-known outflow sources \citep[e.g.,][]{blake95}, but are slightly shifted (up to $\pm$5 km\,s$^{-1}$) with respect to a $\varv_{\rm lsr}$ of 8.5 km\,s$^{-1}$. This indicates that the flows may be caused by J-type shocks rather than C-type shocks \citep{hollenbach97}. Molecular emission is expected to have a Gaussian velocity profile around the shock velocity in a J-type shock, whereas emission caused by C-type shocks is expected to show a "classic", triangular line profile. However, since the observed lines are only shifted by $\pm$5 km\,s$^{-1}$ at most, so the J-type shock speed would be lower than this. CH$_3$OH is also efficiently destroyed in the sputtering process in J-type shocks at velocities greater than 10 km\,s$^{-1}$. The actual CH$_3$OH enhancement may be lower in the outflows in Serpens compared to e.g., L1157, something which is also discussed for the case of NGC2071 \citep{garay00}. Another possibility is that the flows are C-type shocks moving very close to the plane of the sky, in which case the line-wings will not be prominent. This is suported by SiO $J=2-1$ observations of SMM4, where the lines were also found to be narrow \citep[$FWHM\sim$5 km\,s$^{-1}$;][]{garay02}. With the present data-set it is not possible to differentiate the two scenarios.

Emission was integrated over the velocity intervals given in
\citet{davis99}. From this the CO column density is calculated
under the assumption that emission is optically thin and that the gas
temperature is 100 K. The calculation of the column density is
insensitive to the choice of temperature, as long as it is in the
interval of $\sim$ 50--150 K. The calculation is more sensitive to the
choice of velocity interval over which CO emission is integrated, as
this may vary by up to a factor of 2, when changing the velocity
interval with 1 km\,s$^{-1}$. The column density is then converted to
total gas column density using a standard abundance ratio of CO:H$_2$
of 10$^{-4}$:1 and results are compared to CH$_3$OH column densities
to obtain an abundance. The integrated CO intensities, column
densities and CH$_3$OH abundances are all tabulated in Table
\ref{tab:abun_out}. The gas column density in all three outflow
positions is $\sim$ 10$^{20}$ cm$^{-2}$ resulting in abundances of
$\sim$ 1--10$\times$10$^{-6}$ with respect to H$_2$. Due to the
uncertainties on the derived CH$_3$OH column densities and on the gas
column density, the abundances
should be seen as order of magnitude estimates. Compared to the
envelope abundances of SMM1 and SMM4, this translates to methanol
enhancements in the outflows of $\gtrsim$ 10$^{3}$. This
is similar to what is found in other outflows; for example, the
enhancement for the L1157 blue lobe is of the order of $\sim$ 400
\citep{bachiller95, bachiller97}. It should be noted, however, that
the methanol line profiles are highly asymemtric in the L1157 outflow
as are the CO line profiles.

C$^{18}$O is often used as a tracer of quiescent gas column
density. If C$^{18}$O $J$=3--2 emission is used rather than $^{12}$CO
emission as above, the gas column density is increased by $\sim$ two
orders of magnitude to $\sim$ 10$^{22}$~cm$^{-2}$. This leads to a
decrease in CH$_3$OH abundance and therefore also a decrease in the
enhancement. Representing the other extreme, typical
enhancements are of the order of 5--50 (see Table
\ref{tab:abun_out}). Ideally,
  emission should be integrated over the same velocity interval for
  comparison, but in this case, part of the CH$_3$OH line overlaps
  with the $^{12}$CO profile, part of it overlaps with the C$^{18}$O
  profile, which is why these two abundance measurements should be
  seen as extreme values.

\begin{table*}
\caption{Gas column density, $N$(H$_2$) and CH$_3$OH abundance in
  outflow knots as determined from $^{12}$CO $J$=3--2 and C$^{18}$O
  $J$=3--2 emission.}
\label{tab:abun_out}
\center\begin{tabular}{l r r r r|r r r r}
\hline\hline
 & \multicolumn{4}{c|}{$^{12}$CO} & \multicolumn{4}{c}{C$^{18}$O} \\
 & $\int T_{\rm MB}$ d$\varv$ & $N_{\rm CO}^{\rm of}$ &
$x$(CH$_3$OH) & $x_{\rm of}/x_{\rm env}$ & $\int T_{\rm MB}$
d$\varv$ & $N_{\rm CO}$ & $x$(CH$_3$OH) & $x_{\rm of}/x_{\rm
  env}$ \\
Source & (K\,km\,s$^{-1}$) & (10$^{16}$ cm$^{-2}$) &
$\times$10$^{-6}$ & & (K\,km\,s$^{-1}$) & (10$^{18}$ cm$^{-2}$) &
$\times$10$^{-8}$ \\
\hline
SMM1-S & 27.5 & 1.2 & 5.1  & 5000 & 3.6 & 1.1 & 5.2 & 50 \\
SMM4-W & 40.7 & 0.9 & 5.0  & 1000 & 4.6 & 1.4 & 3.1 &  6 \\
SMM4-S & 44.7 & 1.7 & 19.6 & 3900 & 9.8 & 2.9 & 12  & 25 \\
\hline
\end{tabular}
\end{table*}

\section{Discussion}\label{sect:dis}

\begin{figure*}
\center
\includegraphics[width=1.\columnwidth]{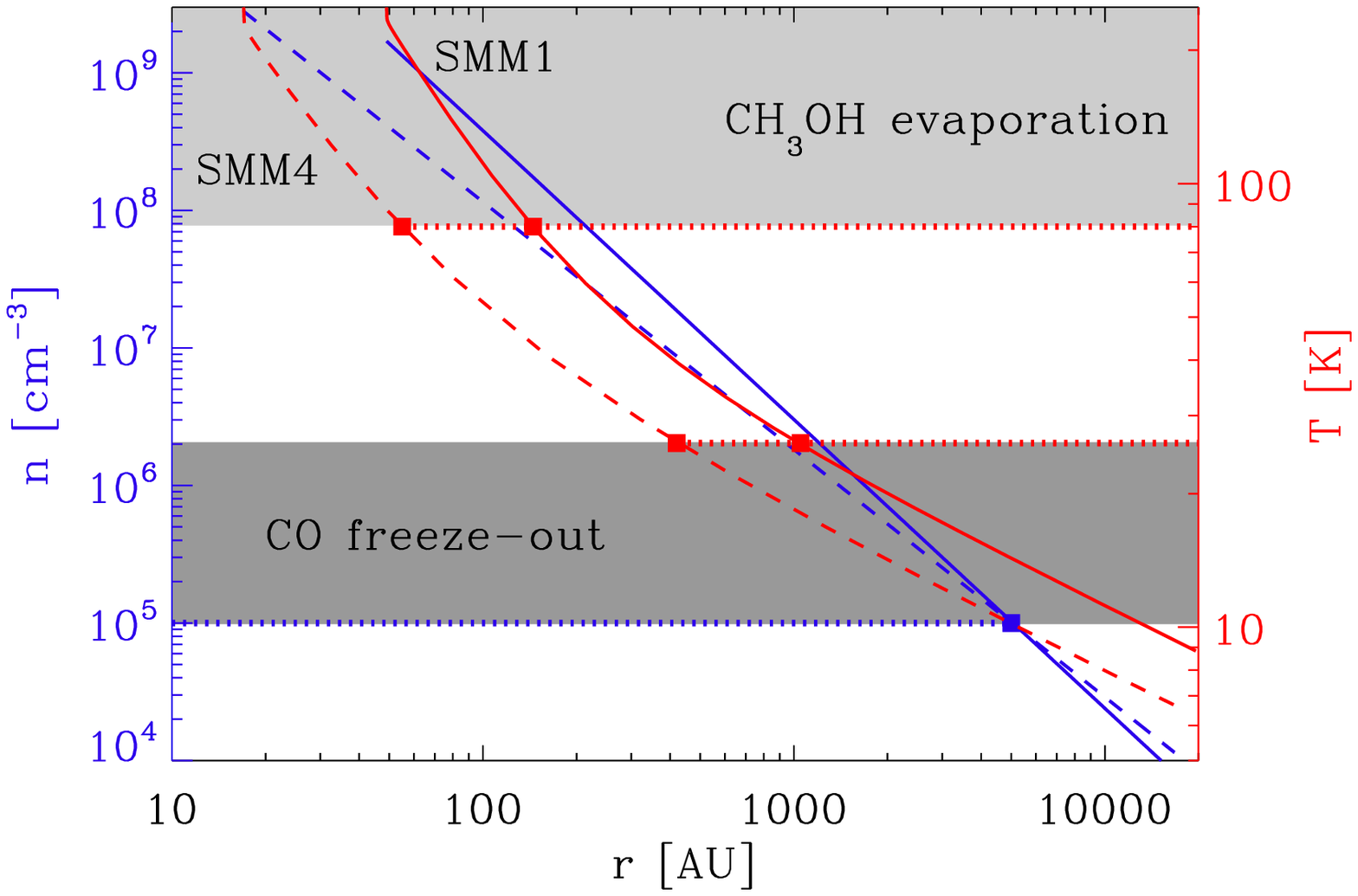}
\includegraphics[width=0.5\columnwidth]{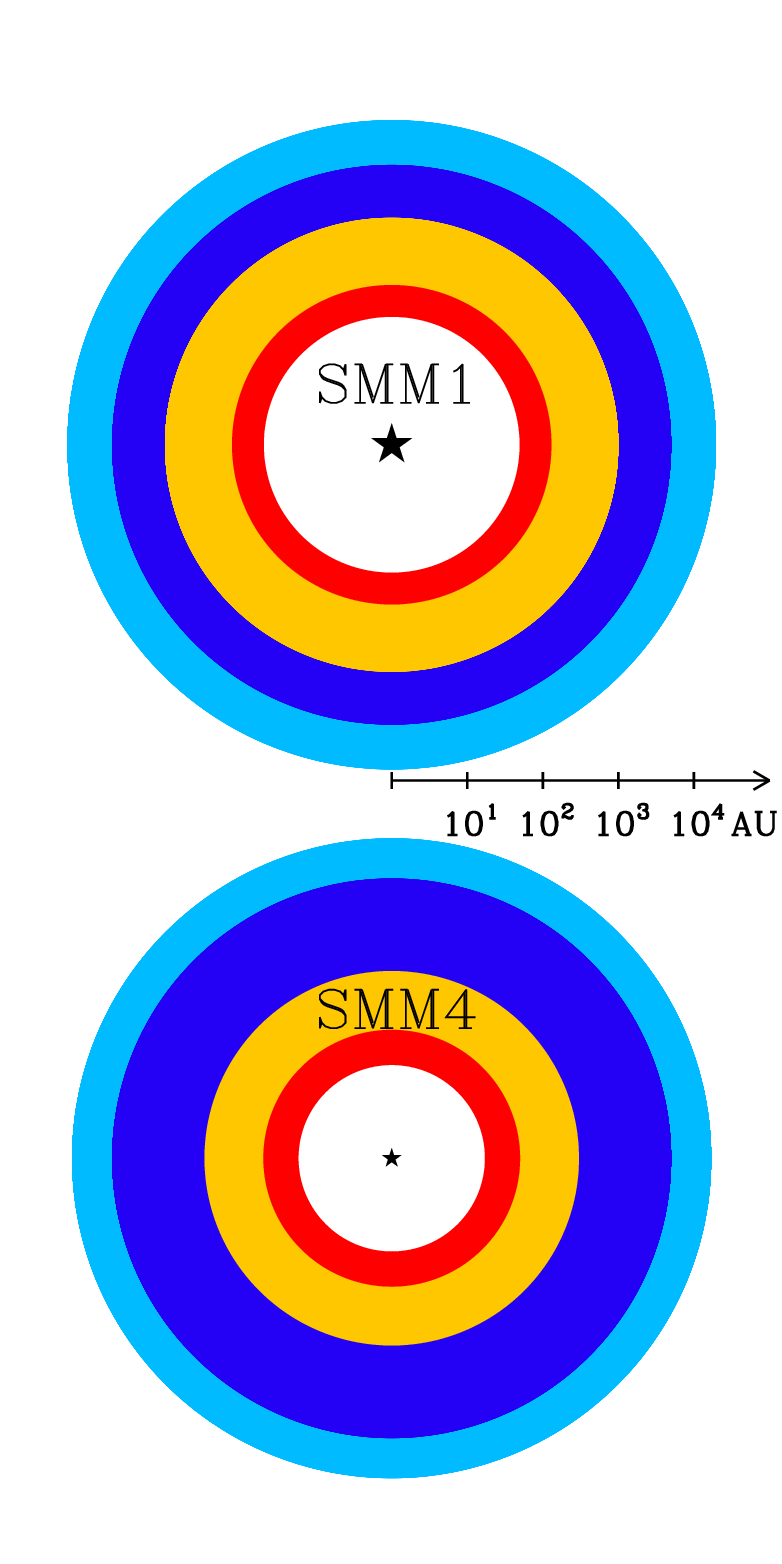}
\caption{Density and temperature profiles of the SMM1 and SMM4
  envelopes as obtained from {\sc Dusty} modelling. The radial density
  profile is displayed in blue (left axis) and kinetic dust
  temperature in red (right axis). The profiles are shown for the SMM1
  envelope (full line) and the SMM4 envelope (dashed line). The CO
  freeze-out zone, characterised by temperatures lower than 25~K and
  densities greater than 10$^5$ cm$^{-3}$, is shown in dark gray,
  whereas the methanol evaporation zone ($T$ $>$ 80~K) is shown in
  light gray. To the right is a cartoon illustrating the differences
  between the envelopes surrounding SMM1 and SMM4. Both illustrations
  are to scale and the dark blue regions correspond to the CO
  freeze-out zones while red indicates CH$_3$OH evaporation
  zones. Note the much smaller CO freeze-out zone for SMM1.}
\label{fig:model}
\end{figure*}

The inferred abundances are all larger than what can be produced by
pure gas-phase reactions \citep[e.g.][]{garrod06}. Thus the observed
CH$_3$OH gas must be formed on grains and subsequently have desorbed. In
the following, possible scenarios for methanol desorption and
excitation are discussed. Finally speculations on the nature of the
individual young stellar objects in Serpens are made.

\subsection{CH$_3$OH grain surface formation}

Methanol is formed on the surfaces of interstellar dust grains through
hydrogenation of CO, a mechanism which has been studied in detail in
the laboratory and in theoretical models of surface chemistry
\citep{hiraoka02, watanabe02, fuchs09}. To form large amounts of
CH$_3$OH, it is imperative that CO freezes effectively out onto
(water-ice covered) dust grains. Toward several pre-stellar cores
\citep[e.g., B68,][]{bergin02} and Class 0 objects
\citep[e.g.,][]{jorgensen05b} CO has been observed to freeze out very
efficiently at temperatures lower than $\sim$~20~K and densities
greater than $\sim$ 10$^5$ cm$^{-3}$. This ``catastrophic'' CO
freeze-out has been observed directly through observations of CO ice
abundances which show an increase in the amount of solid CO with
respect to H$_2$O ice in the densest regions
\citep{pontoppidan06}. Through use of {\sc Dusty} modelling 
discussed above it is possible to estimate the extent of the CO
freeze-out zone in each of the Serpens Class 0 objects quantitatively,
which is illustrated in Fig. \ref{fig:model}. Here the predicted
density and temperature profiles are shown as a function of distance
from the protostar itself. The density profiles of all four envelopes
are very similar, however SMM1 stands out in terms of temperature
profile. Due to the higher luminosity (30~$L_\odot$) of SMM1 with
respect to the other source luminosities ($\sim$~5~$L_\odot$), the
envelope temperature is also higher throughout.

The zone over which CO freezes out in SMM1 is significantly smaller than
that of the other sources, SMM3, SMM4 and S68N. In the case of SMM1,
the present freeze-out zone extends from $\sim$ 1000--5000 AU whereas
the other envelopes have freeze-out zones starting at a few hundred AU
and extend out to the same distance as for SMM1 (see
Fig. \ref{fig:model}). This indicates that at present CO is not
freezing out efficiently in the SMM1 envelope, so any efficient
methanol formation has probably ceased at this point in time. The
colder envelopes surrounding SMM3, SMM4 and S68N could still be
forming CH$_3$OH.

In terms of observing methanol directly in the ice itself,
\citet{pontoppidan04} studied a small region extending south of
SMM4 from 4000 AU to 12\,000 AU by observing the 3.54~$\mu$m CH$_3$OH
features in absorption against background and embedded stars. The
extent is illustrated by the
white line in Fig. \ref{fig:find}. In this region the CH$_3$OH-ice
abundance is constant at 28\% with respect to water ice or 3$\times
10^{-5}$ with respect to gas-phase H$_2$. This corroborates the
interpretation presented here, that the gas phase abundance of
CH$_3$OH is low and constant out to 12\,000 AU in the SMM4 envelope. Beyond
this line the CH$_3$OH-ice abundance drops by at least an order of
magnitude and \citeauthor{pontoppidan04} could only determine upper
limits. It is interesting to note that the location where the ice
abundance drops is where one of the outflow knots starts (SMM4-S; see
Fig. \ref{fig:find}). Thus the reason for the drop is a
combination of the envelope being more tenuous ($n_{\rm H}<$
10$^4$ cm$^{-3}$) far from the protostar, so that CO does not
freeze out very efficiently, while at the same time whatever methanol
is in the ice is sputtered into the gas phase by the outflow.

\citet{cuppen09} have recently studied the formation of
CH$_3$OH on ice surfaces using a Monte Carlo method, in which the
gas-grain chemistry based on laboratory data is simulated
microscopically over long time-scales. The limiting factors in
producing methanol is the availability of both CO and H on the grain
surfaces, and so results show that CO hydrogenates efficiently to form
CH$_3$OH, especially at temperatures lower than 12 K where atomic
hydrogen can be retained efficiently. After 10$^{5}$ years CH$_3$OH
may form up to 100 individual mono-layers on the grain, comparable to
that of water ice. Hence, the ice abundance of methanol in the outer
parts of the envelope will depend strongly on the temperature. Because
the envelopes of SMM3, SMM4 and S68N are significantly colder than
that of SMM1, they would still be actively producing methanol, which
could explain the higher gas phase abundances (see Table \ref{tab:abun}).

\begin{figure}
\center\includegraphics[width=0.8\columnwidth]{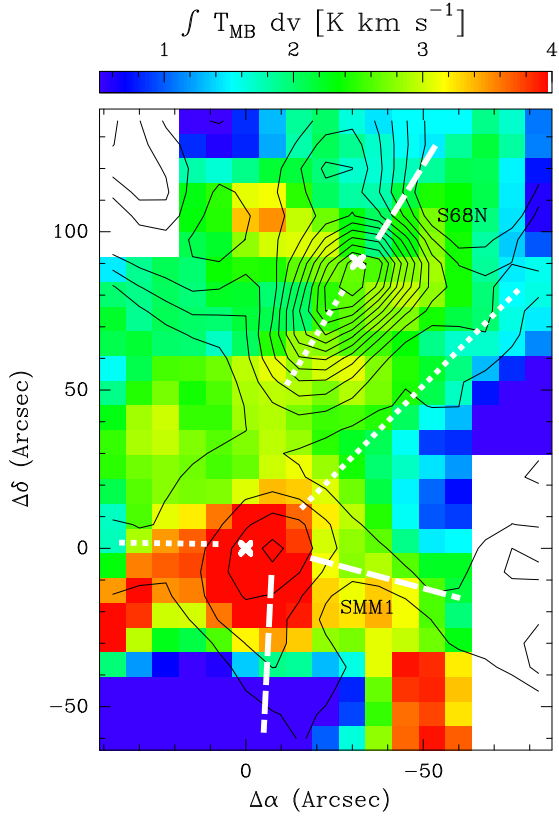}
\caption{Integrated C$^{18}$O, $J$=3--2 emission ($\int T_{\rm mb}
  {\rm d}\varv$) is shown in colour, with integrated CH$_3$OH
  7$_0$--6$_0$ A$^+$ emission overlaid as contours. The image is
  centered on Serpens SMM1 with S68N located to the north. Outflow
  directions are indicated by dashed (red-shifted) and dotted
  (blue-shifted) lines.}
\label{fig:co_ch3oh}
\end{figure}

Observations show a different behaviour of C$^{18}$O at the center
position of S68N compared with the center position of SMM1.
Figure \ref{fig:co_ch3oh} presents C$^{18}$O $J$=3--2 emission 
in colour and methanol emission from the 7$_0$--6$_0$ A$^+$
line overlaid as contours. The morphology of the C$^{18}$O emission
shows a peak toward SMM1, as expected for a warm envelope, and a ring
of emission toward S68N. The
C$^{18}$O abundances towards the two central positions are measured to
be 3 and 1$\times$10$^{-8}$ respectively, based on integrated C$^{18}$O
intensities of 4.3 and 2.4 K\,km\,s$^{-1}$ and the gas column
densities derived from dust continuum emission (Table
\ref{tab:abun}). This is to be compared to a standard C$^{18}$O
abundance of 1.8$\times$10$^{-7}$ for a normal abundance of CO/H$_2$ =
10$^{-4}$ and an $^{16}$O/$^{18}$O ratio of 550
  \citep{wilson94}. Thus, CO is depleted
by a factor of $\sim$ 6 and 18 for the two sources as averaged
  over the entire envelope. For these two sources there is a clear
anti-correlation between CO and CH$_3$OH gas phase abundances.

\begin{figure}
\center\includegraphics[width=\columnwidth]{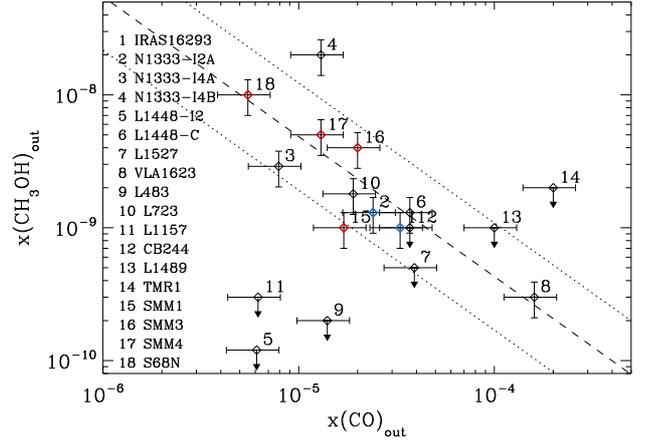}
\caption{CH$_3$OH gas abundance versus CO gas abundance for a selection of
  Class 0 and I sources based on data and analysis from
  \citet{schoier02, maret05, jorgensen05} and this work. The Serpens
  sources are in red and IRAS16293-2422 and
  NGC1333-IRAS2A are in blue. Sources for which 
  only upper methanol limits exist are marked by arrows. The two
  dashed lines indicate the 1$\sigma$ confidence limits of the
  power-law fit.}
\label{fig:c18o_ch3oh}
\end{figure}

To examine whether this anti-correlation is unique to Serpens or a
general feature of embedded sources, the outer-envelope CH$_3$OH
abundances of the sample of \citet{maret05} and 
\citet{jorgensen05} was coupled with CO abundances for the same
sources from \citet{jorgensen02} (see Fig. \ref{fig:c18o_ch3oh}). The
CO gas phase abundance averaged over the extent of the envelope is
taken as a tracer of CO depletion in
the sense that the total gas and ice CO abundance is assumed to be
constant. For sources where the presence of a jump zone can
reproduce the observed CH$_3$OH emission, only the outer abundance is
plotted here. A typical uncertainty of 30\% is assumed both for the CO
and CH$_3$OH abundances. Except for three sources (L1157, L1448-I2 and
L483) there is a correlation between the two abundances as
illustrated by the value of the Pearson correlation coefficient of $r$
= 0.70. 
The Serpens sources are characterized by the combination of low
CO-abundance and high CH$_3$OH abundance. This result implies that the
CH$_3$OH gas phase abundance is directly related to the current
production of CH$_3$OH in the outer, cold parts of the envelope, and
that any difference from source to source is due to a difference in
the amount of CO frozen out onto the grains. Thus, not all solid CH$_3$OH
is formed during the cold pre-stellar core phase, consistent with the
lack of detected CH$_3$OH ice toward background stars behind quiescent
dense clouds. This also implies that the lack of CH$_3$OH emission at
normal cloud positions between the sources is because the density is
lower, and hence the timescale for CO to freeze out is much higher,
i.e., the absolute CH$_3$OH abundance will drop.

\subsection{CH$_3$OH desorption mechanism}

Once methanol has formed on a grain surface it can desorb according to
two different mechanisms, thermal and non-thermal desorption. In the
first mechanism the entire grain is heated thermally (macroscopic
grain heating), and the icy mantle evaporates entirely, releasing all
adsorbed species into the gas phase. The ice mantles typically
evaporate at grain temperatures of 30--100 K, depending on
species. The other mechanism is non-thermal desorption, in which the
ice mantle is ``heated'' on a microscopic (local) scale, either due to
absorption of a single UV-photon \citep{oberg09b}, impact of a cosmic
ray particle \citep{leger85, hasegawa93, herbst06}, or the binding energy being
released from the formation of a new molecule \citep{garrod07} or
sputtering in outflows \citep{jimenezserra08}.

\subsubsection{Thermal desorption}

In the case of methanol, the thermal evaporation temperature has been
determined experimentally to 80--100 K \citep{brown07, green09} as is also
indicated in Fig. \ref{fig:model}. The lower temperature corresponds
to evaporation of a pure CH$_3$OH ice and the higher to
a mix of CH$_3$OH and H$_2$O. Because CO freezes out on top of a water ice, CH$_3$OH is expected to be present in the same layer. Thus the desorption temperature should be close to that of a pure CH$_3$OH ice ($\sim$ 85 K), which is slightly lower, but not much, than that of CH$_3$OH mixed with H$_2$O. This is also found to be the case experimentally \citep{bisschop07}. Thus, for the thermal-desorption
mechanism to be active, it is necessary that all grains are heated to
greater than 80 K, i.e., close to the protostar itself. From the
physical structure models above it is predicted that the radius at
which $T_{\rm dust} > 80$ K is of the order of 50--100 AU or
0\farcs2--0\farcs4 for $d$=230 pc, depending on source, leading to a beam-dilution factor of
1500--5000. Methanol emission originating from very close to the
protostar itself has previously been observed for two low-mass
sources, IRAS16293-2422 \citep[$d$=125 pc;][]{vandishoeck95,
  schoier02} and NGC1333 IRAS2A \citep[$d$=250 pc;][]{maret05}. Even
though the regions have not been spatially resolved with single-dish
data, it has been
possible to infer the existence of ``hot cores'' based on observations
of high-$K$ lines ($K >$ 3) and measured rotational temperatures of
$>$ 80 K. This indicates that the emission arises in a compact, warm
and dense region, consistent with it originating close to the
protostar itself and also consistent with high-spatial resolution
  interferometer data \citep[e.g.,][]{jorgensen05c, jorgensen07}. For
the sources presented here, it has only been
possible to provide an upper limit on the inner abundance, which in
most cases is a few $\times$10$^{-7}$ with respect to H$_2$. One
notable exception is SMM1, where the upper limit on the inner
abundance is close to the outer abundance, a result that will be
discussed further below (Sect. \ref{sect:hotcore}).

\subsubsection{Non-thermal desorption}

Since the temperature in the outer envelope is significantly lower
than 100 K, the primary desorption mechanism must be non-thermal. This is
supported by the fact that the abundance only follows the column
density, and does not appear to depend on temperature. No attempt is
made to distinguish between the above mentioned mechanisms
here. If desorption is induced by secondary UV photons from cosmic ray
ionization of H$_2$ the gas phase abundance is expected to be
10$^{-4}$--10$^{-3}$ times the ice abundance \citep{oberg09a,
  oberg09c}. In the case of SMM4 where the CH$_3$OH-ice abundance is
$\sim$ 3$\times$10$^{-5}$ with respect to H$_2$ \citep{pontoppidan04}
the UV-induced desorption alone can account for a gas phase abundance
of 10$^{-9}$--10$^{-8}$. This is remarkably close to the observed gas
phase abundance of $\sim$ 2$\times$10$^{-8}$.

\citet{garrod07} modelled the release of CH$_3$OH from the grain
surface into the gas phase by examining whether the release of binding
energy would be enough to evaporate the molecule into the gas phase. 
They found that the fraction was probably in the range of 1--10\%, but
could not pin it down any further. In the case of SMM4, results presented
here combined with ice observations indicate that much less than 1--10\%
is desorbed, since the gas/ice abundance is $\sim$ 10$^{-4}$.
Results presented in \citet{hasegawa93} show that direct cosmic ray
desorption is not an efficient desorption mechanism for tightly bound
species like H$_2$O and CH$_3$OH when compared to
the accretion timescale. In fact, they find that the difference
between the two rates is of the order of 10$^4$. Previous studies show
that direct cosmic ray induced desorption is most efficient for
volatile species \citep[see, e.g.,][]{shen04,roberts07} and for
species adsorbed on very small grains \citep[$r \leq 0.05~\mu$m;][]{herbst06}.

\subsubsection{Shock-induced desorption}\label{sect:shock_des}

In outflows the main desorption mechanism is sputtering of the grain
mantle, i.e., impacts of high-temperature gas molecules and atoms,
primarily H$_2$ and He. This can in principle be compared to the direct
cosmic ray desorption mechanism, except that the flux of H$_2$ and He
is orders of magnitude higher and the impact energy is much lower. The
efficiency of this desorption mechanism is clearly seen from the fifth
column of Table \ref{tab:abun_out}, where the CH$_3$OH abundance in
the outflow positions is compared to the ambient envelope
abundance. In particular, the SMM4-S
methanol outflow abundance is $\sim$ 10$^{-5}$ with respect to H$_2$, i.e.,
only a factor of 3 lower than the measured CH$_3$OH-ice abundance in a
region located $\sim$ 15$''$ (4000 AU) to the north \citep[see
Fig. \ref{fig:find} and][]{pontoppidan04}. If the envelope ice-abundance
in this region were also 3$\times$10$^{-5}$ prior to the impact of the
shock, this would indicate that one third or more of the CH$_3$OH ice
is sputtered from the grain mantle. However, the measured gas-phase
abundance as determined from $^{12}$CO emission is an upper limit as
briefly discussed above. The true value of the abundance probably lies
in the range of 10$^{-8}$ to 10$^{-6}$ where the lower limit is
obtained from C$^{18}$O emission.

\subsection{The nature of the embedded YSOs in Serpens}\label{sect:hotcore}

Two low-mass YSOs have been confirmed observationally to have the same
chemical characteristics as high-mass hot cores,
IRAS16293-2422 and NGC1333-IRAS2A. These two sources are at the same evolutionary stage as SMM1 based on classical indicators, e.g., $T_{\rm bol}$. The characteristics 
are (1) a large number of detected, saturated, complex organic
molecules and (2) a high rotational temperature for methanol ($>$ 80 K).
At the same time, neither of these two envelopes show signs of
extended methanol emission \citep[][Kristensen et al. in
  prep.]{vandishoeck95}. IRAS16293-2422 is a close binary system
surrounded by a circum-binary envelope. The inner parts of this
envelope are passively heated to temperatures $>$ 80 K, which, along
with small-scale shocks impinging on the inner wall of the envelope
\citep{chandler05}, give rise to the observed hot-core
signature. NGC1333-IRAS2 is also a binary (possibly triple) system,
but the projected separation is greater ($\sim$ 30\arcsec). It is
believed that the same mechanisms are at play in NGC1333-IRAS2A
causing the hot-core signatures \citep{maret05, jorgensen05}.
In the following the Serpens sources, and in particular SMM1, will be
compared to these two low-mass hot cores, and differences will be
discussed.

There are several possible explanations for the lack of ``hot core''
characteristics in the Serpens sources which may be categorized in the
following manner:
\begin{enumerate}
\item Physical: No gas is present close to the source, or no hot
  gas is present. 
\item Chemical: Hot gas is present, but the methanol abundance is
  very low.
\item Observational: Warm methanol is present close to the
  source, but either the extent is very small, or emission is
  optically thick.
\end{enumerate}

\citet{choi09} recently reported that SMM1 is a
close binary with a projected separation of $\sim$ 500 AU. The primary
would be the sub-mm source associated with SMM1 (SMM1a), while the
binary (SMM1b) would be less embedded. SMM1b can be associated with a
source observed by {\it Spitzer} at wavelengths shorter than 24 $\mu$m,
where SMM1a is not detected. If the projected distance is similar to
the actual separation between the two sources, then this could have
cleared the inner part of the envelope of gas, explaining the absence
of hot gas close to the (sub-mm) source. However, both IRAS16293-2422
and NGC1333-IRAS2A have been shown to have inner holes or cavities
similar in size to that of SMM1 \citep{schoier02, jorgensen05b},
therefore this cannot be the entire explanation.

Recent millimeter interferometry of SMM1 indicates that the region
close to the protostar has been cleared of gas. Both \citet{enoch09} and
\citet{vankempen09} report the detection of a resolved disk surrounding
SMM1. The disk is unusually massive ($\sim$ 1 $M_\odot$) and has a
modelled radius of up to 300 AU \citep{enoch09}, while the inner 500
AU of the envelope have been cleared of gas. In such a dense disk,
only the upper-most layers will be warm or hot, depending on the
distance to the protostar, as the dust extinction grows rapidly towards
the midplane of the disk. Thus, the column density of hot methanol
will be very low. Through SED modeling of the disk, \citet{enoch09}
estimate that the inclination is 15\degr, i.e., nearly
face-on. This is in conflict with observations of the 3.6-cm radio
jet, which indicate that the outflow is moving very close to the plane
of the sky \citep{rodriguez89, moscadelli06}. Moreover, the $^{12}$CO line
profiles are very symmetric along the large-scale outflow, i.e., the molecular
outflows are also moving close to the plane of the sky. This would
indicate that the disk is seen close to edge-on. If so, then
the beam-dilution will be very high implying that any hot part
will not be detectable in our single-dish beam. The disk surrounding
NGC1333-IRAS2A is less massive compared to SMM1 \citep[$\sim$ 0.056
  $M_\odot$;][]{jorgensen09} and is viewed closer to face-on resulting
in less extinction and lower beam dilution. This implies that the
high-$K$ emission observed in IRAS16293-2422 and NGC1333-IRAS2A is
originating in the disk close to the protostar. The actual heating
mechanism (passive heating or small-scale shocks) cannot be
distinguished with the current observations.

If the inner part of the envelope has not been cleared out of warm gas
or if the disk is seen face-on, then it is possible that the abundance
of methanol is low. As has been shown by current observations, the
CH$_3$OH gas-phase abundance is comparatively high in the outer parts of the
envelopes, so to decrease the abundance in the inner part of the
system, destruction of CH$_3$OH must be present. This destruction
mechanism must be very efficient if it is to destroy all CH$_3$OH in
the hot-core parts of the envelope, where the abundance is expected to
rise to $>$ 10$^{-6}$ w.r.t. H$_2$. CH$_3$OH can be
destroyed in the gas phase through direct reactions with other
species, however the rate coefficients are typically low. An
alternative mechanism of both the gas and the ice destruction is
UV-dissociation. However, this destruction
mechanism must also be at play in IRAS16293-2422 and
NGC1333-IRAS2A and there is currently no reason why
UV-photodissociation would be more efficient in Serpens sources
than in the other two nor that the UV-field is enhanced in SMM1.

Finally, a low-mass hot core may be present in all of the Serpens sources,
but not observable. This can be due to beam dilution or due to optical
depth effects. The region over which methanol desorbs from the grain
surface is expected to be $\sim$ 100 AU at most, and so beam
dilution would be of the order of 10$^3$. However, the same beam
dilution would apply to at least NGC1333-IRAS2A, which is located at a
similar distance of 250 pc, and can thus not
be used as an argument. The line optical depth has been calculated in the
{\sc Ratran} simulations, and is typically of the order of 0.1 or
less for the transitions observed here, even for lines arising in the
inner-most part of the envelope. The dust opacity at 338 GHz is less
than 0.08 at all times, as estimated from the {\sc Dusty} modelling.

Of the three explanations presented above the first is the more
plausible if the distance is indeed 230 pc as assumed. Recent VLBA observations indicate that the distance may be closer to 415 pc \citep{dzib10}, in which case the third explanation is more plausible. The other two reasons can be disproved through comparison
with IRAS16293-2422 and NGC1333-IRAS2A. However, only SMM1 has been
suggested to have a massive disk, and it is also the source that
resembles the two 
low-mass hot cores the most in terms of luminosity. The other sources
(SMM3, SMM4 and S68N) have considerably lower luminosities by factors
4--10. Thus it may be that the hot-core regions around these sources
are indeed much smaller and beam-diluted ($\sim$ 5$\times$ 10$^3$),
comparable to several of the low-mass sources in the sample of
\citet{maret05} and \citet{jorgensen05}. With the current observations
only upper limits have been determined of the molecular
abundance in the inner envelope of the lower-luminosity sources.

\section{Summary}\label{sect:conc}

Maps of rotationally excited methanol in the Serpens Molecular Core
have been presented. Emission arises from the molecular envelopes of
four deeply embedded sources and their associated outflows. In
particular, three outflow knots have been identified based on their
strong methanol emission. Mapping shows methanol emission in all of
the four envelopes to be extended out to ranges of $\sim$~10\,000~AU.
The abundance is constant in the outer parts of the envelope
with a value of 10$^{-9}$--10$^{-8}$, depending on source. The
methanol abundance at outflow positions is enhanced by up to 2--3 orders of
magnitude with respect to the ambient abundance.
The measured envelope abundances are consistent with non-thermal
desorption of solid-state methanol, through, for example,
UV-photodesorption. At outflow positions the enhanced abundance may be
explained by sputtering of the grain mantles. The symmetric, slightly shifted line profiles point to the outflows being either caused by J-type shocks or moving along the plane of the sky.

The CO gas abundance is found to be anti-correlated with the methanol
gas abundance in the Serpens maps. This result has been extended to
literature data of a large sample of other, similarly deeply
embedded sources. The reason is that the more CO is frozen out from the
gas phase, the higher the CO ice abundance, and this adsorbed CO is then
converted to CH$_3$OH ice on the grain surface from where it desorbs
non-thermally. The non-thermal desorption mechanism implies that the
gas-phase abundance follows the solid-state abundance closely,
something which has directly shown to be the case here. Thus, the
differences in abundance between the four sources can be directly
related to the current production of CH$_3$OH and reflects how much CO
is frozen out now rather than in some possible colder past.

The Serpens sources do not contain the chemical signatures of low-mass hot
cores such as IRAS16293-2422 and NGC1333-IRAS2A. In the case of SMM1,
the more luminous of the Serpens sources and resembling most closely
the two low-mass hot cores, the most likely explanation is that it
harbours a massive disk which is seen close to edge-on. The other
three Serpens sources are all lower in mass and luminosity, and here
the absence of a hot-core signature could be ascribed to beam
dilution, as also holds for the majority of the deeply embedded
sources.

In conclusion, the CH$_3$OH emission is found to trace the following:
(1) energetic input into cold gas, primarily through outflow
interaction; (2) column density of cold gas in the outer envelope; (3)
reactions and subsequent desorption of grain surface products. 

Because water closely resembles methanol in the sense that it is
exclusively formed on grains and it desorbs from grain surfaces
at $T>$ 100 K, it is expected that the abundance structure will
resemble that of methanol. This is to be tested with upcoming
water-observations of these sources as part of the Herschel Guaranteed
Time Key Project ``Water in Star-forming Regions with Herschel''
(WISH).

\begin{acknowledgements}
Astrochemistry at Leiden Observatory is supported by a Spinoza prize
and by NWO grant 614.041.004. The authors would like to thank the
staff at the JCMT for technical help. Floris van der Tak is thanked
for help with updating the LAMDA database and Karin {\"O}berg for very
stimulating disucssions. TvK is grateful to the SMA for supporting his
research at the CfA.
\end{acknowledgements}

\bibliographystyle{aa}
\bibliography{bibliography}

\begin{thebibliography}{81}
\expandafter\ifx\csname natexlab\endcsname\relax\def\natexlab#1{#1}\fi

\bibitem[{{Bachiller} {et~al.}(1998){Bachiller}, {Codella}, {Colomer},
  {Liechti}, \& {Walmsley}}]{bachiller98}
{Bachiller}, R., {Codella}, C., {Colomer}, F., {Liechti}, S., \& {Walmsley},
  C.~M. 1998, \aap, 335, 266

\bibitem[{{Bachiller} {et~al.}(1995){Bachiller}, {Liechti}, {Walmsley}, \&
  {Colomer}}]{bachiller95}
{Bachiller}, R., {Liechti}, S., {Walmsley}, C.~M., \& {Colomer}, F. 1995, \aap,
  295, 51

\bibitem[{{Bachiller} \& {P\'erez Guti\'errez}(1997)}]{bachiller97}
{Bachiller}, R. \& {P\'erez Guti\'errez}, M. 1997, \apjl, 487, 93

\bibitem[{{Bachiller} \& {Tafalla}(1999)}]{bachiller99}
{Bachiller}, R. \& {Tafalla}, M. 1999, in NATO ASIC Proc. 540: The Origin of
  Stars and Planetary Systems, ed. {C.~J.~Lada \& N.~D.~Kylafis}, 227--+

\bibitem[{{Bergin} {et~al.}(2002){Bergin}, {Alves}, {Huard}, \&
  {Lada}}]{bergin02}
{Bergin}, E.~A., {Alves}, J., {Huard}, T., \& {Lada}, C.~J. 2002, \apjl, 570,
  L101

\bibitem[{{Bisschop}(2007)}]{bisschop07}
{Bisschop}, S.~E. 2007, PhD thesis, Leiden Observatory, Leiden University,
  P.O.~Box 9513, 2300 RA Leiden, The Netherlands

\bibitem[{{Blake} {et~al.}(1995){Blake}, {Sandell}, {van Dishoeck},
  {Groesbeck}, {Mundy}, \& {Aspin}}]{blake95}
{Blake}, G.~A., {Sandell}, G., {van Dishoeck}, E.~F., {et~al.} 1995, \apj, 441,
  689

\bibitem[{{Brown} \& {Bolina}(2007)}]{brown07}
{Brown}, W.~A. \& {Bolina}, A.~S. 2007, \mnras, 374, 1006

\bibitem[{{Casali} {et~al.}(1993){Casali}, {Eiroa}, \& {Duncan}}]{casali93}
{Casali}, M.~M., {Eiroa}, C., \& {Duncan}, W.~D. 1993, \aap, 275, 195

\bibitem[{{Ceccarelli} {et~al.}(2000){Ceccarelli}, {Loinard}, {Castets},
  {Tielens}, \& {Caux}}]{ceccarelli00}
{Ceccarelli}, C., {Loinard}, L., {Castets}, A., {Tielens}, A.~G.~G.~M., \&
  {Caux}, E. 2000, \aap, 357, L9

\bibitem[{{Chandler} {et~al.}(2005){Chandler}, {Brogan}, {Shirley}, \&
  {Loinard}}]{chandler05}
{Chandler}, C.~J., {Brogan}, C.~L., {Shirley}, Y.~L., \& {Loinard}, L. 2005,
  \apj, 632, 371

\bibitem[{{Choi}(2009)}]{choi09}
{Choi}, M. 2009, \apj, 705, 1730

\bibitem[{{Cuppen} {et~al.}(2009){Cuppen}, {van Dishoeck}, {Herbst}, \&
  {Tielens}}]{cuppen09}
{Cuppen}, H.~M., {van Dishoeck}, E.~F., {Herbst}, E., \& {Tielens}, A.~G.~G.~M.
  2009, \aap, 508, 275

\bibitem[{{Dartois} {et~al.}(1999){Dartois}, {Schutte}, {Geballe}, {Demyk},
  {Ehrenfreund}, \& {D'Hendecourt}}]{dartois99}
{Dartois}, E., {Schutte}, W., {Geballe}, T.~R., {et~al.} 1999, \aap, 342, L32

\bibitem[{{Davis} {et~al.}(1999){Davis}, {Matthews}, {Ray}, {Dent}, \&
  {Richer}}]{davis99}
{Davis}, C.~J., {Matthews}, H.~E., {Ray}, T.~P., {Dent}, W.~R.~F., \& {Richer},
  J.~S. 1999, \mnras, 309, 141

\bibitem[{{Di Francesco} {et~al.}(2008){Di Francesco}, {Johnstone}, {Kirk},
  {MacKenzie}, \& {Ledwosinska}}]{difrancesco08}
{Di Francesco}, J., {Johnstone}, D., {Kirk}, H., {MacKenzie}, T., \&
  {Ledwosinska}, E. 2008, \apjs, 175, 277

\bibitem[{{Dzib} {et~al.}(2010){Dzib}, {Loinard}, {Mioduszewski}, {Boden},
  {Rodriguez}, \& {Torres}}]{dzib10}
{Dzib}, S., {Loinard}, L., {Mioduszewski}, A.~J., {et~al.} 2010, ArXiv e-prints

\bibitem[{{Eiroa} {et~al.}(2008){Eiroa}, {Djupvik}, \& {Casali}}]{eiroa08}
{Eiroa}, C., {Djupvik}, A.~A., \& {Casali}, M.~M. 2008, in Handbook of Star
  Forming Regions, Volume II: The Southern Sky ASP Monograph Publications,
  Vol.~5., ed. B.~{Reipurth}, 693

\bibitem[{{Enoch} {et~al.}(2009){Enoch}, {Corder}, {Dunham}, \&
  {Duch{\^e}ne}}]{enoch09}
{Enoch}, M.~L., {Corder}, S., {Dunham}, M.~M., \& {Duch{\^e}ne}, G. 2009, \apj,
  707, 103

\bibitem[{{Enoch} {et~al.}(2007){Enoch}, {Glenn}, {Evans}, {Sargent}, {Young},
  \& {Huard}}]{enoch07}
{Enoch}, M.~L., {Glenn}, J., {Evans}, II, N.~J., {et~al.} 2007, \apj, 666, 982

\bibitem[{{Evans} {et~al.}(2009){Evans}, {Dunham}, {J{\o}rgensen}, {Enoch},
  {Mer{\'{\i}}n}, {van Dishoeck}, {Alcal{\'a}}, {Myers}, {Stapelfeldt},
  {Huard}, {Allen}, {Harvey}, {van Kempen}, {Blake}, {Koerner}, {Mundy},
  {Padgett}, \& {Sargent}}]{evans09}
{Evans}, N.~J., {Dunham}, M.~M., {J{\o}rgensen}, J.~K., {et~al.} 2009, \apjs,
  181, 321

\bibitem[{{Friberg} {et~al.}(1988){Friberg}, {Hjalmarson}, {Madden}, \&
  {Irvine}}]{friberg88}
{Friberg}, P., {Hjalmarson}, A., {Madden}, S.~C., \& {Irvine}, W.~M. 1988,
  \aap, 195, 281

\bibitem[{{Froebrich}(2005)}]{froebrich05}
{Froebrich}, D. 2005, \apjs, 156, 169

\bibitem[{{Fuchs} {et~al.}(2009){Fuchs}, {Cuppen}, {Ioppolo}, {Romanzin},
  {Bisschop}, {Andersson}, {van Dishoeck}, \& {Linnartz}}]{fuchs09}
{Fuchs}, G.~W., {Cuppen}, H.~M., {Ioppolo}, S., {et~al.} 2009, \aap, 505, 629

\bibitem[{{Garay} {et~al.}(2000){Garay}, {Mardones}, \&
  {Rodr{\'{\i}}guez}}]{garay00}
{Garay}, G., {Mardones}, D., \& {Rodr{\'{\i}}guez}, L.~F. 2000, \apj, 545, 861

\bibitem[{{Garay} {et~al.}(2002){Garay}, {Mardones}, {Rodr{\'{\i}}guez},
  {Caselli}, \& {Bourke}}]{garay02}
{Garay}, G., {Mardones}, D., {Rodr{\'{\i}}guez}, L.~F., {Caselli}, P., \&
  {Bourke}, T.~L. 2002, \apj, 567, 980

\bibitem[{{Garrod} {et~al.}(2006){Garrod}, {Park}, {Caselli}, \&
  {Herbst}}]{garrod06}
{Garrod}, R., {Park}, I.~H., {Caselli}, P., \& {Herbst}, E. 2006, in Faraday
  Discussions 133, 51

\bibitem[{{Garrod} {et~al.}(2007){Garrod}, {Wakelam}, \& {Herbst}}]{garrod07}
{Garrod}, R.~T., {Wakelam}, V., \& {Herbst}, E. 2007, \aap, 467, 1103

\bibitem[{{Gibb} {et~al.}(2004){Gibb}, {Whittet}, {Boogert}, \&
  {Tielens}}]{gibb04}
{Gibb}, E.~L., {Whittet}, D.~C.~B., {Boogert}, A.~C.~A., \& {Tielens},
  A.~G.~G.~M. 2004, \apjs, 151, 35

\bibitem[{{Green} {et~al.}(2009){Green}, {Bolina}, {Chen}, {Collings}, {Brown},
  \& {McCoustra}}]{green09}
{Green}, S.~D., {Bolina}, A.~S., {Chen}, R., {et~al.} 2009, \mnras, 398, 357

\bibitem[{{Harvey} {et~al.}(2007){Harvey}, {Mer{\'{\i}}n}, {Huard}, {Rebull},
  {Chapman}, {Evans}, \& {Myers}}]{harvey07}
{Harvey}, P., {Mer{\'{\i}}n}, B., {Huard}, T.~L., {et~al.} 2007, \apj, 663,
  1149

\bibitem[{{Hasegawa} \& {Herbst}(1993)}]{hasegawa93}
{Hasegawa}, T.~I. \& {Herbst}, E. 1993, \mnras, 261, 83

\bibitem[{{Herbst} \& {Cuppen}(2006)}]{herbst06}
{Herbst}, E. \& {Cuppen}, H.~M. 2006, Proceedings of the National Academy of
  Science, 103, 12257

\bibitem[{{Hiraoka} {et~al.}(2002){Hiraoka}, {Sato}, {Sato}, {Sogoshi},
  {Yokoyama}, {Takashima}, \& {Kitagawa}}]{hiraoka02}
{Hiraoka}, K., {Sato}, T., {Sato}, S., {et~al.} 2002, \apj, 577, 265

\bibitem[{{Hogerheijde} \& {van der Tak}(2000)}]{hogerheijde00}
{Hogerheijde}, M.~R. \& {van der Tak}, F.~F.~S. 2000, \aap, 362, 697

\bibitem[{{Hogerheijde} {et~al.}(1999){Hogerheijde}, {van Dishoeck},
  {Salverda}, \& {Blake}}]{hogerheijde99}
{Hogerheijde}, M.~R., {van Dishoeck}, E.~F., {Salverda}, J.~M., \& {Blake},
  G.~A. 1999, \apj, 513, 350

\bibitem[{{Hollenbach}(1997)}]{hollenbach97}
{Hollenbach}, D. 1997, in IAU Symposium, Vol. 182, Herbig-Haro Flows and the
  Birth of Stars, ed. {B.~Reipurth \& C.~Bertout}, 181--198

\bibitem[{{Hurt} \& {Barsony}(1996)}]{hurt96}
{Hurt}, R.~L. \& {Barsony}, M. 1996, \apjl, 460, L45+

\bibitem[{{Ivezic} \& {Elitzur}(1997)}]{ivezic97}
{Ivezic}, Z. \& {Elitzur}, M. 1997, \mnras, 287, 799

\bibitem[{{Jim{\'e}nez-Serra} {et~al.}(2008){Jim{\'e}nez-Serra}, {Caselli},
  {Mart{\'{\i}}n-Pintado}, \& {Hartquist}}]{jimenezserra08}
{Jim{\'e}nez-Serra}, I., {Caselli}, P., {Mart{\'{\i}}n-Pintado}, J., \&
  {Hartquist}, T.~W. 2008, \aap, 482, 549

\bibitem[{{J{\o}rgensen} {et~al.}(2007){J{\o}rgensen}, {Bourke}, {Myers}, {Di
  Francesco}, {van Dishoeck}, {Lee}, {Ohashi}, {Sch{\"o}ier}, {Takakuwa},
  {Wilner}, \& {Zhang}}]{jorgensen07}
{J{\o}rgensen}, J.~K., {Bourke}, T.~L., {Myers}, P.~C., {et~al.} 2007, \apj,
  659, 479

\bibitem[{{J{\o}rgensen} {et~al.}(2005{\natexlab{a}}){J{\o}rgensen}, {Bourke},
  {Myers}, {Sch{\"o}ier}, {van Dishoeck}, \& {Wilner}}]{jorgensen05c}
{J{\o}rgensen}, J.~K., {Bourke}, T.~L., {Myers}, P.~C., {et~al.}
  2005{\natexlab{a}}, \apj, 632, 973

\bibitem[{{J{\o}rgensen} {et~al.}(2002){J{\o}rgensen}, {Sch{\"o}ier}, \& {van
  Dishoeck}}]{jorgensen02}
{J{\o}rgensen}, J.~K., {Sch{\"o}ier}, F.~L., \& {van Dishoeck}, E.~F. 2002,
  \aap, 389, 908

\bibitem[{{J{\o}rgensen} {et~al.}(2004){J{\o}rgensen}, {Sch{\"o}ier}, \& {van
  Dishoeck}}]{jorgensen04}
{J{\o}rgensen}, J.~K., {Sch{\"o}ier}, F.~L., \& {van Dishoeck}, E.~F. 2004,
  \aap, 416, 603

\bibitem[{{J{\o}rgensen} {et~al.}(2005{\natexlab{b}}){J{\o}rgensen},
  {Sch{\"o}ier}, \& {van Dishoeck}}]{jorgensen05}
{J{\o}rgensen}, J.~K., {Sch{\"o}ier}, F.~L., \& {van Dishoeck}, E.~F.
  2005{\natexlab{b}}, \aap, 437, 501

\bibitem[{{J{\o}rgensen} {et~al.}(2005{\natexlab{c}}){J{\o}rgensen},
  {Sch{\"o}ier}, \& {van Dishoeck}}]{jorgensen05b}
{J{\o}rgensen}, J.~K., {Sch{\"o}ier}, F.~L., \& {van Dishoeck}, E.~F.
  2005{\natexlab{c}}, \aap, 435, 177

\bibitem[{{J{\o}rgensen} {et~al.}(2009){J{\o}rgensen}, {van Dishoeck},
  {Visser}, {Bourke}, {Wilner}, {Lommen}, {Hogerheijde}, \&
  {Myers}}]{jorgensen09}
{J{\o}rgensen}, J.~K., {van Dishoeck}, E.~F., {Visser}, R., {et~al.} 2009,
  \aap, 507, 861

\bibitem[{{Larsson} {et~al.}(2000){Larsson}, {Liseau}, {Men'shchikov},
  {Olofsson}, {Caux}, {Ceccarelli}, {Lorenzetti}, {Molinari}, {Nisini},
  {Nordh}, {Saraceno}, {Sibille}, {Spinoglio}, \& {White}}]{larsson00}
{Larsson}, B., {Liseau}, R., {Men'shchikov}, A.~B., {et~al.} 2000, \aap, 363,
  253

\bibitem[{{L\'eger} {et~al.}(1985){L\'eger}, {Jura}, \& {Omont}}]{leger85}
{L\'eger}, A., {Jura}, M., \& {Omont}, A. 1985, \aap, 144, 147

\bibitem[{{Leurini} {et~al.}(2007){Leurini}, {Schilke}, {Wyrowski}, \&
  {Menten}}]{leurini07}
{Leurini}, S., {Schilke}, P., {Wyrowski}, F., \& {Menten}, K.~M. 2007, \aap,
  466, 215

\bibitem[{{Maret} {et~al.}(2005){Maret}, {Ceccarelli}, {Tielens}, {Caux},
  {Lefloch}, {Faure}, {Castets}, \& {Flower}}]{maret05}
{Maret}, S., {Ceccarelli}, C., {Tielens}, A.~G.~G.~M., {et~al.} 2005, \aap,
  442, 527

\bibitem[{{McMullin} {et~al.}(2000){McMullin}, {Mundy}, {Blake}, {Wilking},
  {Mangum}, \& {Latter}}]{mcmullin00}
{McMullin}, J.~P., {Mundy}, L.~G., {Blake}, G.~A., {et~al.} 2000, \apj, 536,
  845

\bibitem[{{McMullin} {et~al.}(1994){McMullin}, {Mundy}, {Wilking}, {Hezel}, \&
  {Blake}}]{mcmullin94}
{McMullin}, J.~P., {Mundy}, L.~G., {Wilking}, B.~A., {Hezel}, T., \& {Blake},
  G.~A. 1994, \apj, 424, 222

\bibitem[{{Moscadelli} {et~al.}(2006){Moscadelli}, {Testi}, {Furuya}, {Goddi},
  {Claussen}, {Kitamura}, \& {Wootten}}]{moscadelli06}
{Moscadelli}, L., {Testi}, L., {Furuya}, R.~S., {et~al.} 2006, \aap, 446, 985

\bibitem[{{M{\"u}ller} {et~al.}(2001){M{\"u}ller}, {Thorwirth}, {Roth}, \&
  {Winnewisser}}]{muller01}
{M{\"u}ller}, H.~S.~P., {Thorwirth}, S., {Roth}, D.~A., \& {Winnewisser}, G.
  2001, \aap, 370, L49

\bibitem[{{{\"O}berg} {et~al.}(2009{\natexlab{a}}){{\"O}berg}, {Bottinelli}, \&
  {van Dishoeck}}]{oberg09a}
{{\"O}berg}, K.~I., {Bottinelli}, S., \& {van Dishoeck}, E.~F.
  2009{\natexlab{a}}, \aap, 494, L13

\bibitem[{{{\"O}berg} {et~al.}(2009{\natexlab{b}}){{\"O}berg}, {Garrod}, {van
  Dishoeck}, \& {Linnartz}}]{oberg09c}
{{\"O}berg}, K.~I., {Garrod}, R.~T., {van Dishoeck}, E.~F., \& {Linnartz}, H.
  2009{\natexlab{b}}, \aap, 504, 891

\bibitem[{{{\"O}berg} {et~al.}(2009{\natexlab{c}}){{\"O}berg}, {van Dishoeck},
  \& {Linnartz}}]{oberg09b}
{{\"O}berg}, K.~I., {van Dishoeck}, E.~F., \& {Linnartz}, H.
  2009{\natexlab{c}}, \aap, 496, 281

\bibitem[{{Ossenkopf} \& {Henning}(1994)}]{ossenkopf94}
{Ossenkopf}, V. \& {Henning}, T. 1994, \aap, 291, 943

\bibitem[{{Pontoppidan}(2006)}]{pontoppidan06}
{Pontoppidan}, K.~M. 2006, \aap, 453, L47

\bibitem[{{Pontoppidan} {et~al.}(2008){Pontoppidan}, {Boogert}, {Fraser}, {van
  Dishoeck}, {Blake}, {Lahuis}, {{\"O}berg}, {Evans}, \&
  {Salyk}}]{pontoppidan08}
{Pontoppidan}, K.~M., {Boogert}, A.~C.~A., {Fraser}, H.~J., {et~al.} 2008,
  \apj, 678, 1005

\bibitem[{{Pontoppidan} {et~al.}(2004){Pontoppidan}, {van Dishoeck}, \&
  {Dartois}}]{pontoppidan04}
{Pontoppidan}, K.~M., {van Dishoeck}, E.~F., \& {Dartois}, E. 2004, \aap, 426,
  925

\bibitem[{{Pottage} {et~al.}(2004){Pottage}, {Flower}, \& {Davis}}]{pottage04}
{Pottage}, J.~T., {Flower}, D.~R., \& {Davis}, S.~L. 2004, \mnras, 352, 39

\bibitem[{{Roberts} {et~al.}(2007){Roberts}, {Rawlings}, {Viti}, \&
  {Williams}}]{roberts07}
{Roberts}, J.~F., {Rawlings}, J.~M.~C., {Viti}, S., \& {Williams}, D.~A. 2007,
  \mnras, 382, 733

\bibitem[{{Rodriguez} {et~al.}(1989){Rodriguez}, {Curiel}, {Moran}, {Mirabel},
  {Roth}, \& {Garay}}]{rodriguez89}
{Rodriguez}, L.~F., {Curiel}, S., {Moran}, J.~M., {et~al.} 1989, \apjl, 346,
  L85

\bibitem[{{Schnee} {et~al.}(2005){Schnee}, {Ridge}, {Goodman}, \&
  {Li}}]{schnee05}
{Schnee}, S.~L., {Ridge}, N.~A., {Goodman}, A.~A., \& {Li}, J.~G. 2005, \apj,
  634, 442

\bibitem[{{Sch{\"o}ier} {et~al.}(2002){Sch{\"o}ier}, {J{\o}rgensen}, {van
  Dishoeck}, \& {Blake}}]{schoier02}
{Sch{\"o}ier}, F.~L., {J{\o}rgensen}, J.~K., {van Dishoeck}, E.~F., \& {Blake},
  G.~A. 2002, \aap, 390, 1001

\bibitem[{{Sch{\"o}ier} {et~al.}(2005){Sch{\"o}ier}, {van der Tak}, {van
  Dishoeck}, \& {Black}}]{schoier05}
{Sch{\"o}ier}, F.~L., {van der Tak}, F.~F.~S., {van Dishoeck}, E.~F., \&
  {Black}, J.~H. 2005, \aap, 432, 369

\bibitem[{{Shen} {et~al.}(2004){Shen}, {Greenberg}, {Schutte}, \& {van
  Dishoeck}}]{shen04}
{Shen}, C.~J., {Greenberg}, J.~M., {Schutte}, W.~A., \& {van Dishoeck}, E.~F.
  2004, \aap, 415, 203

\bibitem[{{Smith} {et~al.}(2003){Smith}, {Hills}, {Withington}, {Richer},
  {Leech}, {Williamson}, {Gibson}, {Dace}, {Ananthasubramanian}, {Barker},
  {Baldwin}, {Stevenson}, {Doherty}, {Molloy}, {Quy}, {Lush}, {Hales}, {Dent},
  {Pain}, {Wall}, {Hastings}, {Graham}, {Baillie}, {Laidlaw}, {Bennett},
  {Laidlaw}, {Duncan}, {Ellis}, {Redman}, {Wooff}, {Yeung}, {Fitzsimmons},
  {Avery}, {Derdall}, {Josephson}, {Anthony}, {Atwal}, {Chylek}, {Shutt},
  {Friberg}, {Rees}, {Philips}, {Kroug}, {Klapwijk}, \& {Zijlstra}}]{smith03}
{Smith}, H., {Hills}, R.~E., {Withington}, S., {et~al.} 2003, in Society of
  Photo-Optical Instrumentation Engineers (SPIE) Conference, Vol. 4855, ., ed.
  T.~G. {Phillips} \& J.~{Zmuidzinas}, 338--348

\bibitem[{{Tafalla} {et~al.}(2000){Tafalla}, {Myers}, {Mardones}, \&
  {Bachiller}}]{tafalla00}
{Tafalla}, M., {Myers}, P.~C., {Mardones}, D., \& {Bachiller}, R. 2000, \aap,
  359, 967

\bibitem[{{Testi} \& {Sargent}(1998)}]{testi98}
{Testi}, L. \& {Sargent}, A.~I. 1998, \apjl, 508, L91

\bibitem[{{van der Tak} {et~al.}(2000){van der Tak}, {van Dishoeck}, \&
  {Caselli}}]{vandertak00}
{van der Tak}, F.~F.~S., {van Dishoeck}, E.~F., \& {Caselli}, P. 2000, \aap,
  361, 327

\bibitem[{{van Dishoeck} \& {Blake}(1998)}]{vandishoeck98}
{van Dishoeck}, E.~F. \& {Blake}, G.~A. 1998, \araa, 36, 317

\bibitem[{{van Dishoeck} {et~al.}(1995){van Dishoeck}, {Blake}, {Jansen}, \&
  {Groesbeck}}]{vandishoeck95}
{van Dishoeck}, E.~F., {Blake}, G.~A., {Jansen}, D.~J., \& {Groesbeck}, T.~D.
  1995, \apj, 447, 760

\bibitem[{{van Kempen} {et~al.}(2009){van Kempen}, {Wilner}, \&
  {Gurwell}}]{vankempen09}
{van Kempen}, T.~A., {Wilner}, D., \& {Gurwell}, M. 2009, \apjl, 706, L22

\bibitem[{{Watanabe} \& {Kouchi}(2002)}]{watanabe02}
{Watanabe}, N. \& {Kouchi}, A. 2002, \apjl, 571, L173

\bibitem[{{White} {et~al.}(1995){White}, {Casali}, \& {Eiroa}}]{white95}
{White}, G.~J., {Casali}, M.~M., \& {Eiroa}, C. 1995, \aap, 298, 594

\bibitem[{{Williams} \& {Myers}(2000)}]{williams00}
{Williams}, J.~P. \& {Myers}, P.~C. 2000, \apj, 537, 891

\bibitem[{{Wilson} \& {Rood}(1994)}]{wilson94}
{Wilson}, T.~L. \& {Rood}, R. 1994, \araa, 32, 191

\bibitem[{{Wolf-Chase} {et~al.}(1998){Wolf-Chase}, {Barsony}, {Wootten},
  {Ward-Thompson}, {Lowrance}, {Kastner}, \& {McMullin}}]{wolf-chase98}
{Wolf-Chase}, G.~A., {Barsony}, M., {Wootten}, H.~A., {et~al.} 1998, \apjl,
  501, L193+

\end{thebibliography}

\appendix
\section{SED data points}

\begin{table*}
\caption{SED points used for {\sc Dusty} modelling.}
\label{tab:sed}
\begin{tabular}{r r r c r r r c r r r c r r r c}
\hline\hline
\multicolumn{4}{c}{SMM1} & \multicolumn{4}{c}{SMM3} &
\multicolumn{4}{c}{SMM4} & \multicolumn{4}{c}{S68N} \\
\hline
$\lambda$ & $F$ & Beam & Ref. & $\lambda$ & $F$ & Beam & Ref. &
$\lambda$ & $F$ & Beam & Ref. & $\lambda$ & $F$ & Beam & Ref. \\
($\mu$m) & (Jy) & (\arcsec) & & ($\mu$m) & (Jy) & (\arcsec) & &
($\mu$m) & (Jy) & (\arcsec) & & ($\mu$m) & (Jy) & (\arcsec) \\
\hline
70.1 & 313 & 90 & 1 & 24 & 0.109 &  & 6 & 60 & 12 & 60 & 2 & 50 & 16 &32
& 8 \\
73.6 & 353 & 90 & 1 & 60 & 22 & 60 & 2 & 100 & 60 & 60 & 2 & 60 & 8 & 60 & 2 \\
77.3 & 402 & 90 & 1 & 70 & 12.2 &  & 6 & 450 & 10.8 & 8 & 3 & 70 & 14.9 &  &6\\
81.2 & 445 & 90 & 1 & 100 & 70 & 100 & 2 & 800 & 4.13 & 14 & 4 & 70 & 5.664 &  &9\\
85.2 & 489 & 90 & 1 & 450 & 7.1 & 8 & 3 & 850 & 2.9 & 14 & 3 & 100 & 37 & 100 & 2 \\
89.5 & 536 & 90 & 1 & 800 & 2.01 & 14 & 4 & 1100 & 1.47 & 19 & 4 & 450
& 25 & 8 & 10 \\
94.0 & 573 & 90 & 1 & 850 & 1.5 & 14 & 3 & 1300 & 0.92 & 22 & 4 & 450 & 9.1
& 8 & 3 \\
98.7 & 612 & 90 & 1 & 1100 & 1.11 & 19 & 4 & 1400 & 1.108 & 1 & 7 & 850 & 1.6
& 14 & 3 \\
103.6 & 638 & 90 & 1 & 1300 & 0.92 & 22 & 4 & 2000 & 0.6 & 35 & 4 & 850 & 3.3 & 20&10 \\
108.8 & 661 & 90 & 1 & 1400 & 0.651 & 2 & 7 & 2700 & 0.143 & 2 & 7 & 1100 & 2.32 & 40&9 \\
114.2 & 685 & 90 & 1 & 2700 & 0.09 & 4 & 7 & 3200 & 0.097 & 4 & 7 & 1100 & 3.78 & 40 &9 \\
119.9 & 698 & 90 & 1 & 3400 & 0.043 & 5 & 7 & 3400 & 0.075 & 5 & 7 & 1250 & 0.9 & 32 & 8\\
125.9 & 713 & 90 & 1 & & & & & & & &  &  2727 & 0.0361 & 7 & 11 \\
132.2 & 733 & 90 & 1 & & & & & & & &  &  3100 & 0.036 & 32 & 8 \\
138.8 & 760 & 90 & 1 \\
145.8 & 770 & 90 & 1 \\
153.1 & 785 & 90 & 1 \\
160.7 & 785 & 90 & 1 \\
100 & 435 & 100 & 2 \\
450 & 35.7 & 8 & 3 \\
850 & 6.1 & 14 & 3 \\
1100 & 3.47 & 19 & 4 \\
1300 & 2.52 & 22 & 4 \\
2000 & 0.91 & 35 & 4 \\
3000 & 0.233 & 5 & 5 \\ \hline
\end{tabular}
References: (1) \citet{larsson00}, Larsson, priv. comm. (2) \citet{hurt96} (3)
\citet{davis99} (4) \citet{casali93} (5) \citet{testi98} (6)
\citet{harvey07} (8) \citet{hogerheijde99} (8) \citet{mcmullin94} (9)
\citet{enoch07} (10) \citet{wolf-chase98} (11) \citet{williams00}
\end{table*}

\end{document}